\documentclass[aps,preprint,groupedaddress]{revtex4-1}

\usepackage{amsmath}
\usepackage{amssymb}
\usepackage{graphicx}
\usepackage{color}
\usepackage{comment}
\usepackage{subfig}

\makeatletter
\renewcommand{\thetable}{\@arabic\c@table}
\usepackage[justification=raggedright]{caption}
\usepackage{url}
\usepackage{multirow}

\begin{document}

\title{\normalsize\ Workforce migration and its economic implications:\\ A perspective from social media in China}

\author{Xiaoqian Hu$^{1}$, Junjie Wu$^{1,2}$ and Jichang Zhao$^{1,2,\star}$}

\affiliation{
$^1$School of Economics and Management, Beihang University \\
$^2$Beijing Advanced Innovation Center for Big Data and Brain Computing, Beihang University\\
$^\star$Corresponding author: jichang@buaa.edu.cn}

\begin{abstract}
The workforce remains the most basic element of social production, even in modern societies. Its migration, especially for developing economies such as China, plays a strong role in the reallocation of productive resources and offers a promising proxy for understanding socio-economic issues. Nevertheless, due to long cycle, expensive cost and coarse granularity, conventional surveys face challenges in comprehensively profiling it. With the permeation of smart and mobile devices in recent decades, booming social media has essentially broken spatio-temporal constraints and enabled the continuous sensing of the real-time mobility of massive numbers of individuals. In this study, we demonstrate that similar to a natural shock, the Spring Festival culturally drives workforce travel between workplaces and hometowns, and the trajectory footprints from social media therefore open a window with unparalleled richness and fine granularity to explore laws in national-level workforce migration. To understand the core driving forces of workforce migration flux between cities in China, various indicators reflecting the benefits and costs of migration are introduced into our prediction model. We find that urban GDP (gross domestic product) and travel time are two excellent indicators to help make predictions. Diverse migration patterns are then revealed by clustering the trajectories, which give important clues to help understand the different roles of Chinese cities in their own development and in regional economic development. These patterns are further explained as a joint effect of the subjective will to seek personal benefits and the capacity requirements of local labour markets. Our study implies that the non-negligible entanglement between social media and macroeconomic behaviours can be insightful for policymaking in social-economic issues.
\end{abstract}


\maketitle

\section{I\lowercase{ntroduction}}
Given its importance, the workforce remains one of the most basic elements of social production in modern societies, even in the age of the knowledge economy. The workforce's intercity migration within a country, usually between hometowns and workplaces, essentially reflects the reallocation of production resources that is profoundly responsible for the normal operation of the national economy and the promotion of production and operating activities. Accordingly, workforce migration exerts a strong influence on social-economic issues~\cite{Abel_2014,Shen_2012}, to name a few examples, including the balanced development of the regional economy, national transportation design, urban infrastructure planning, policymaking in household registration systems and climate change mitigation~\cite{Shayegh_2017}. Understanding the laws of workforce migration at the national scale will therefore help to solve these issues. 

Although individual movement is stochastic, the collective behaviour of migration is not random but closely linked to extensive social, economic and even political factors~\cite{Castaneda_2017}. Laws or patterns underlying group travel can emerge as the population size increases~\cite{Liu_2014,Chepizhko_2016}, suggesting that the laws of workforce migration can be probed and modelled. In fact, tremendous efforts have been devoted to profiling and forecasting human migration flux in recent years~\cite{Simini_2012,Zagheni_2014,Yan_2014,Liang_2013,Barchiesi_2015,Yxy2017nc}. In addition, diverse patterns of human mobility are revealed~\cite{Levy_2010,Blumenstock_2012,Noulas_2012,Hawelka_2014,Thiemann_2010,Liu_2014,Karemera_2000,Fagiolo_2013, Lee_2014, Ghosh_2017}, and their driving forces are also extensively explored~\cite{Stark_1991,Fan_2005,Ravenstein_1885,Lee_1966,Taylor_1999,Decressin_1995,Carrington_1996,Taylor_2003, Lixiao_2016, Wanglz_2018}. 

However, most existing studies assume that human migration behaviours are determined only by passively accepting an attraction from cities, neglecting people's subjective will to seek benefits and the capability requirement of distinct labour markets that indeed profoundly affect workforce movement. In fact, from the viewpoint of economics, the driving force behind workforce migration is a higher economic benefit or the Pareto optimality of the economy~\cite{Ravenstein_1885,Lee_1966,Decressin_1995,Carrington_1996,Taylor_1999}. The development of science and technology also leads to a major labour market concern that tends to focus more on technical capacity than on the size of the workforce. Moreover, from the cognition perspective, sensitivity regarding influence factors varies across different individuals~\cite{Khasin_2012}. Hence, the cost of migration should not always be represented by the spherical distance, as in previous studies. Richer information should be included as indicators to demonstrate that the workforce possesses different cognitive sensitivities to indicators when making behavioural decisions~\cite{Zhao_2016}.

Apart from the above limitations, the data adopted in previous studies have adverse effects on the ability to carry out an in-depth study of workforce migration. Specifically, survey samples from questionnaires or government censuses constitute the major data sources for previous studies on the workforce~\cite{Blumenstock_2012,Zagheni_2014,Hawelka_2014}. However, it is well known that limited samples, biased sample selection, and the uncontrollable quality of investigations might lead to unreliable survey data. Although government censuses supply valid nationwide records, they involve an enormous expense and are quite time-consuming. Indeed, population censuses usually have a very long survey circle (once per decade, as designated by the government of China), and the resulting non-current investigations are not helpful for flexible and effective policy formulation. Moreover, it is difficult to obtain a fine-grained survey in geographical terms. For instance, the population census in China supplies only the migration flux between provinces (see http://www.stats.gov.cn/tjsj/pcsj/), while fine-grained inter-city trajectories are necessary to investigate workforce migration on the city level. In summary, traditional investigations cannot offer the desired data in terms of cost, granularity and reliability.

Online media, such as Twitter and its variant Weibo in China, have boomed in recent decades and offer an unprecedented opportunity to understand human movements. Massive numbers of users leave their digital footprints as messages on social media, including both spatial and temporal information. The continuous penetration of smart and mobile devices further boosts the scale and granularity of these footprints, facilitating the investigation of human movements on national and systemic levels. In order to distinguish the workforce migration behaviours from the movements for other purposes, we adopt the Spring Festival scenario. Being the most important traditional custom in China, the Spring Festival culturally drives people working in other places back to their hometowns to reunite with their families and celebrate the Chinese New Year. It can be treated as a natural shock in workforce movement on a national scale, leading to the full-scale migration with high traffic peaks before and after the Lunar New Year, namely, the Spring Festival travel rush. Accordingly, the resulting trajectories form an ideal data source for modelling workforce migration on a national scale. 

In this paper, we collect massive trajectories of movements from Weibo during the 2017 Spring Festival travel rush and establish a nearly complete picture of workforce migration in China at the city level. To explore the core driving force of this systematic migration, the most widespread in the world, we introduce economic factors and alternative cost measures to the well-known gravity model (GM) framework~\cite{Zipf_1946, Krings_2009,Lambiotte_2008} to make high-quality predictions. By evaluating the forecasting performance for real-life data, urban GDP and the travel time between cities in flux are two key factors that lead to the best performance. We then investigate the diverse patterns of workforce migration by clustering trajectories using the $K$-means algorithm, and the prediction accuracy of the models improves accordingly. More importantly, two dominant patterns are highlighted, {\it i.e.}, the migration flows from small neighbourhoods to local core cities and the flows between cities in good economic statuses, especially in the eastern coastal region. We further provide a rigorous analysis of gaming between interest articulation and cost elusion in different patterns, and these transfer modes also embody the filter effect of the labour market when the workforce makes migration decisions. The results indeed testify to the interest-driven intrinsic motivation of the workforce and the extrinsic limiting condition of market requirements. Finally, an evaluation approach inspired by the dividing modes is proposed to value regional development in China. Some practical problems, such as the deficiency in regional leading roles for Beijing and Chongqing, are exposed, and a thorough discussion of corresponding strategies is also provided. Our models and findings could shed light on related social-economic issues such as urban planning, workforce management and regional development.  

\section{R\lowercase{esults}}
\subsection{Sensing the trajectories of inter-urban workforce migration}
An increasing number of people in China choose to leave their homes in search of job opportunities because of the relaxation of the policy that had restricted inter-city population movements since the early 1980s. Additionally, a great number of migrating workers travel back to their hometowns during the Spring Festival in what is known as the Spring Festival travel rush~\cite{Li_2016,Wang_2013, Wang_2014} due to its unique cultural and traditional meaning. Lasting 40 days surrounding the Spring Festival, the great travel rush defined by the National Development and Reform Commission begins on the 15th day of the twelfth month of the lunar year and ends on the 25th day of the first lunar month of the following year. We argue that the Spring Festival travel rush can be regarded as a natural experiment on the national scale, and it provides us with reliable crowdsourcing migration trajectories to probe workforce movement patterns~\cite{Xu_2017, Liu_2015}. Combined with the spread of the online social media and mobile devices, we find that it is possible to break through the limitation of traditional methods in order to gain a nearly complete profile on a nationwide scale for such a long period. Note that there must be noise in the collected trajectories, but fortunately, those few noisy movements by non-workforce participants have no pronounced influence on our findings.  

Specifically, messages in Weibo contain the locations where they are posted, and thus, a 4-tuple $(m, u, t, g)$ can be used to denote a message $m$ posted by user $u$ in timestamp $t$ at location $g$ of city-granularity. For a user $u$, an array $L(u) = \{(t_{u,1}, g_{u,1}), (t_{u,2}, g_{u,2})$, \ldots , $(t_{u,m}, g_{u,m})\}$ is defined to denote all locations $u$ experienced, where $t_{u,i}$ represents the time at which $u$ posts a message ($t_{u,1}<t_{u,2}< \ldots<t_{u,m}$) and $g_{u, i}$ is the corresponding posting city. Then, we filter the same locations in which people post messages in subsequent iterations. If $\{(t_{u,i}, g_{u,i}), (t_{u,i+1}, g_{u,i+1}), \ldots, (t_{u, i+k}, g_{t,i+k})\} \subseteq L(u)$, $g_{u, i} = g_{u, i+1}= \ldots = {g_{u, i+k}}$ and $g_{u, i-1} \neq g_{u, i}$ , $ g_{u, i+k+1} \neq g_{u, i+k}$, only $(t_{u, i+k}, g_{u, i+k})$ is reserved. Finally, we obtain $\tilde{L}(u) = \{ (\tilde{t}_{u,1}, \tilde{g}_{u,1}), (\tilde{t}_{u,2}, \tilde{g}_{u,2})$, \ldots , $(\tilde{t}_{u,n}, \tilde{g}_{u,n})\}$, where $\tilde{g}_{u,i} \neq \tilde{g}_{u,i+1}$, $i \in [1, n-1]$. Additionally, the trajectory of $u$ is cut as $(\tilde{g}_{u, 1}, \tilde{g}_{u, 2})$, $(\tilde{g}_{u, 2}, \tilde{g}_{u, 3})$, \ldots, $(\tilde{g}_{u, n-1}, \tilde{g}_{u, n})$. Then, we can gain the transition set by summing all users' trajectories as $\{(g_1, g_2, f_{1,2}), \ldots, (g_i, g_j, f_{i,j}), \ldots\}$, where $f_{i,j}$ denotes the migration flux from city $g_i$ to $g_j$. In the same way, the undirected transition set can be obtained as $\{(g_1, g_2, F_{1,2}), \ldots, (g_i, g_j, F_{i,j}), \ldots\}$, where $F_{i,j}$ represents the number of people in the population transferring between cities $g_i$ and $g_j$, suggesting that $F_{i,j} = f_{i, j} + f_{j, i}$ and $F_{i,j} = F_{j,i}$. More particularly, 371 cities are contained in the transition set, and there are 61,759 city-pairs in undirected trajectories and 120,361 city-pairs in directed trajectories, totaling 41,454,268 migration traces (see Methods). 

To prove the rationality of using data from Weibo to investigate workforce migration problems, we collect national railway line data, 5,878 trains in total, and count the number of trains through city pairs. It shows that the flow of workforce migration has a positive correlation with the number of trains between city pairs (see Fig. S1). Therefore, workforce migration flux acquired from social media is representative and indeed embodies population movement in real life from the perspective of trains. Furthermore, it cannot be neglected that only approximately one-fifth of city pairs extracted from Weibo data can be explained by railway line data, specifically, 12,382 city pairs. According to the comparison above, Weibo reflects richer information about workforce migration than the national railway line does. Therefore, the rationality and completeness of migration data gained from Weibo builds a foundation to carry out the following in-depth investigations.

\subsection{Modelling inter-urban workforce migration}

In the classical gravity model, human migration is assumed to be dependent on the attractiveness and accessibility of locations. The former is always estimated as the population size, and the latter is calculated according to the geographical distance between two locations. In detail, the migration flux between cities $i$ and $j$ is defined as $F_{ij} = a \frac{P_i \cdot P_j}{d_{i j}^ \gamma}$, where $P_i$ denotes the number of people residing in city $i$, $F_{i j}$ represents the migration flow between $i$ and $j$, $d_{i j}$ is the metric of distance between the two cities, $a$ is a constant to adjust the forecast value, and $\gamma$ is usually assumed to be positive. This model implies that the migration flow increases with the population size and decreases with the distance between the two locations. 

The motivation for workforce migration has been extensively investigated in previous efforts, and it has been proven that migrating behaviour is not completely blind or disorderly but rather abides by certain laws. Based on the microeconomic perspective, profit maximization, utility optimization and intuitive revenue enhancement~\cite{Decressin_1995, Carrington_1996,Taylor_1999,Taylor_2003} are treated as the target of supplying labour to discuss the specific influence factors when workers make migration decisions. In addition, from the sociological perspective, human migration aims at improving living conditions and is the result of push and pull dynamics~\cite{Ravenstein_1885, Lee_1966}. More specifically, immigrant areas with abundant job opportunities, higher wage levels and other preferential treatment conditions produce a ``pull" force. Additionally, harsh living conditions in the original place of residence would  ``push" human emigration. Following the idea of a subjective measure for the pursuit of profit maximization, we introduce economic factors rather than demographic factors into our models. Based on this, we propose an extension of GM as follows.
\begin{eqnarray}
\label{eq:gdpGM}
	F_{i j} = a \frac{ E_i^\alpha \cdot E_j^ \beta}{C_{i j}^\gamma},
\end{eqnarray}
where $E_i$ stands for the economic indicator of city $i$, and $C_{i j}$ represents the cost of migration between city $i$ and city $j$. There is an assumption that the labour force would prefer to migrate to cities with a high economic status and more job opportunities, and these cities could provide more potential for the workforce to meet labour market demands, meaning that $\alpha$ and $\beta$ should be positive. Additionally, $\gamma$ is still assumed to be positive, indicating that the workforce prefers to migrate to cities with a low cost of migration.

\begin{figure}[!h]
\includegraphics[width=4.5in]{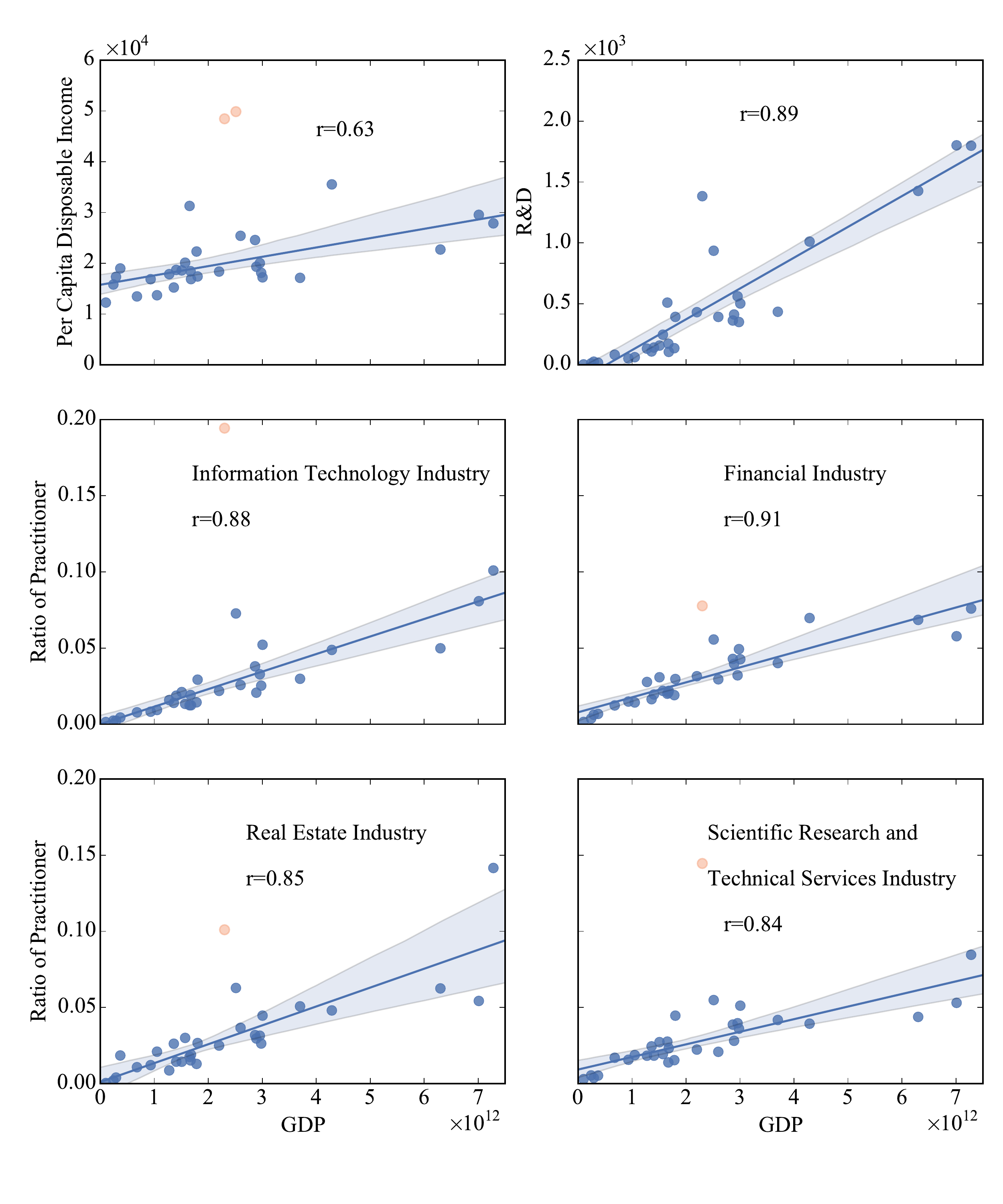}
\caption{
{\bf Relations between GDP and per capita disposable income, R\&D, and talent distribution.}
There is a high correlation between GDP and the per capita disposable income in different provinces. The investment of funds in research and development (R\&D) is positively correlated with GDP in provinces. Meanwhile, the ratios of practitioners in the high-technology industries, such as the information technology industry, financial industry, real estate industry and scientific research and technical services industry, are all significantly positively related to GDP. This finding shows that GDP, as an indicator, has information on potential income, technical workforce levels and labour market requirements. The city with high GDP would have more job opportunities with higher yields, a higher-skilled workforce and high labour requirements. The outliers are marked by orange circles; in detail, Beijing and Shanghai are unusual points in the first subgraphs, and the outliers of the other subgraph denote Beijing.
}
\label{fig:gdp_relation}
\end{figure}

With regard to cities' economic indicators, GDP is one of the most important indices of national economic accounting, containing abundant information about social and economic situations, and it is almost always available, especially in undeveloped areas. As shown in Fig.~\ref{fig:gdp_relation}, the GDP of different provinces is positively correlated with per capita disposable income, implying that the workforce is more likely to have better revenue in a province with high GDP. Furthermore, the investment of funds in research and development (R\&D) has a strong positive correlation with provinces' GDP. Intuitively, this effect is due to greater investment in R\&D and more advanced degrees in the market. Additionally, the ratios of practitioners in high-technology industries, such as the information technology industry, the financial industry, the real estate industry, and the scientific and technical services industry, all show significant positive correlations with GDP in Fig.~\ref{fig:gdp_relation}. This result means that provinces with better economic statuses have higher technology-related workforce requirements and that the market in these provinces could offer more highly skilled labour. As discussed above, GDP can simultaneously reflect a region's income level and required level of labour skills. It is natural to select GDP as an economic indicator in our model, specifically called the G-GM model. Considering the demographic information included in GDP, we select per capita GDP as another economic indicator because it can be better compared to the classical GM model with a population reflecting cities' attraction; we call this model the aveG-GM model. To determine the relationship between the origin and the destination in detail, we introduce a variant of G-GM, named dirG-GM. Unlike the three models introduced above, $E_i$ and $E_j$ in dirG-GM denote the total GDP of the origin and the destination, and $F_{i j}$ in Eq.~\ref{eq:gdpGM} is changed to $f_{i j}$, the migration flux from city $i$ to $j$. Because of the undirected migration flux between cities in G-GM and aveG-GM, we simplify the parameters $\alpha$ and $\beta$ to 1.                 

With respect to the index for migration cost, the migration cost between two cities can be intuitively measured in terms of the geographical distance, or spherical distance; for example, in GM, the spherical distance is calculated using the longitudes and latitudes of city centers. However, due to the complexity of China's geographical circumstances, the same geographical distance does not suggest the same accessibility between cities. For instance, the reachability between two cities in a mountainous area is worse than that in plains regions with the same geographical distance. In this case, direct geographical distance might provide less information to the workforce than metrics such as travel time. Therefore, the wealth of information included in metrics shows great differences. In the meantime, from the perspective of decision making, the sensitivity of the labour force to indicators is diverse. Hence, alternative measures such as the shortest travel distance and travel time, which are both obtained from the navigation system of the Baidu Map, enrich the distance metrics in all models. 
 
\begin{figure}[!h]
\includegraphics[width=7in]{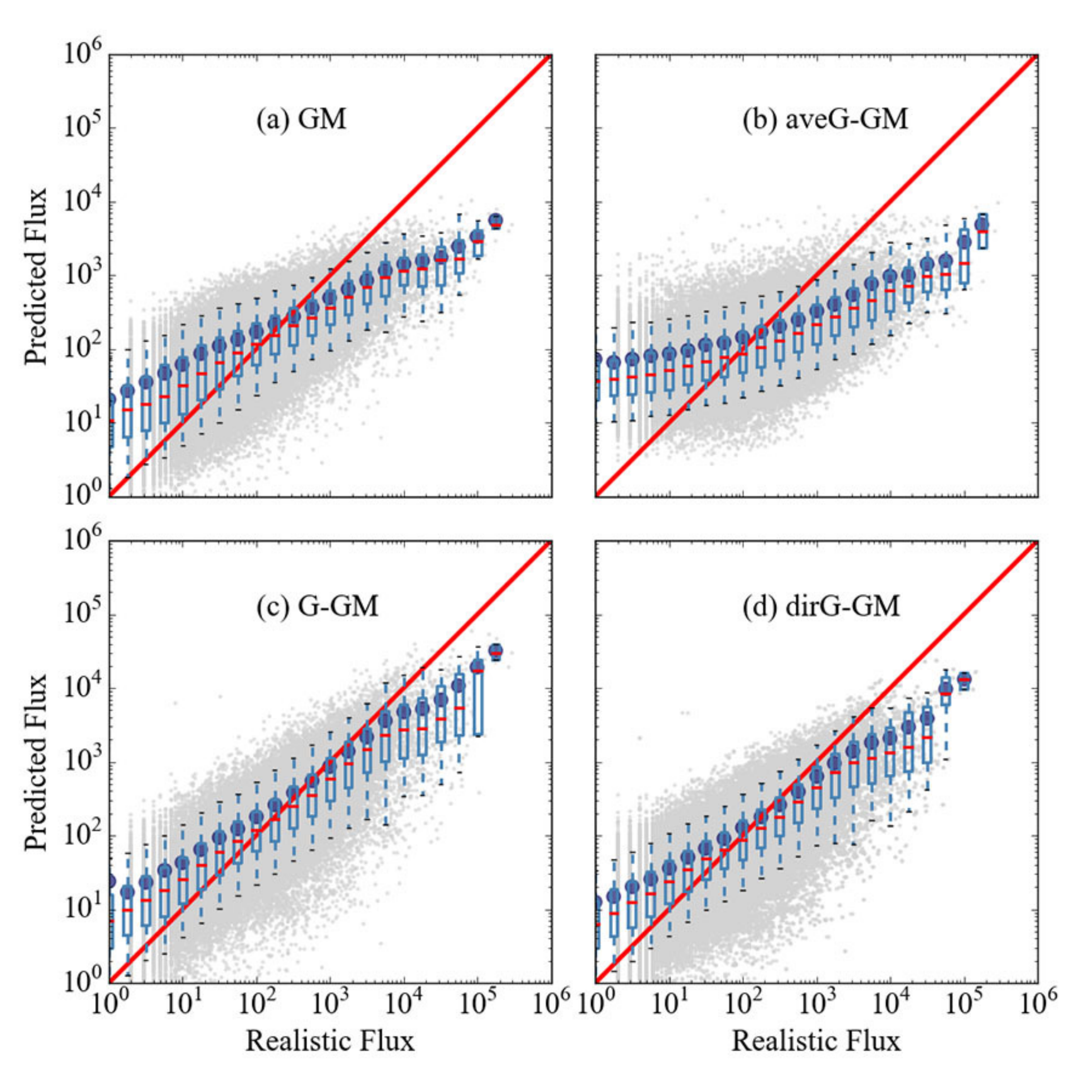}
\caption{
{\bf Modelling flux predictions for four models, with the distance metric being travel time.}
The grey points are a scatter plot of fitness between the actual labour migration flows and forecast flows between each pair of cities. The red lines of $y=x$ indicate the best prediction, and the red points represent the means of the predicted migration flows. Additionally, the whiskers of the box plots show the 5th and 95th percentiles, respectively.
}
\label{fig:model_driving_time}
\end{figure}

\begin{figure}[!h]
\centering
\includegraphics[width=6in]{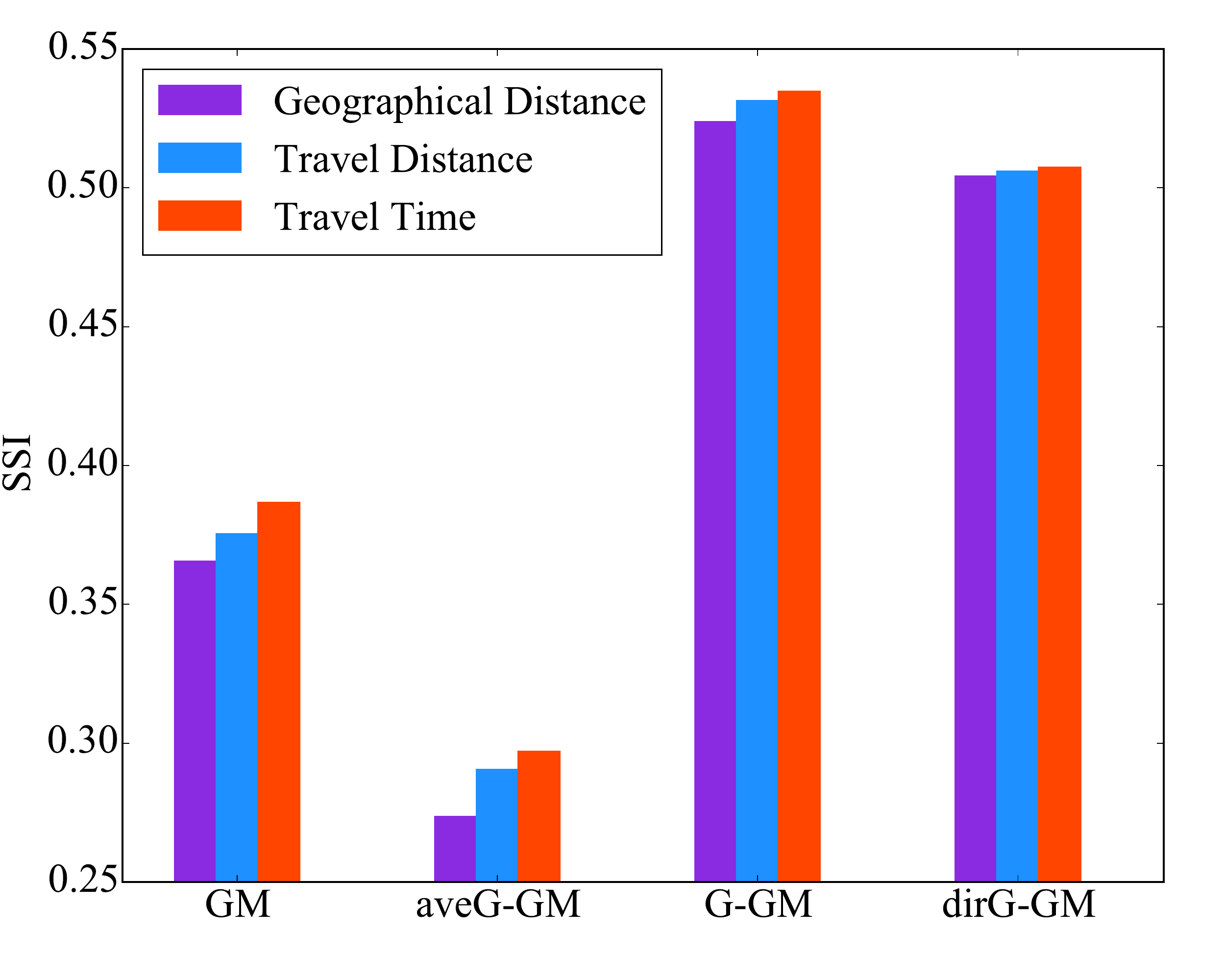}
\caption{
{\bf Comparison of four models based on SSI.}
The bars with different colors show the predictive capability of models with different distance metrics. Higher SSI reflects better accuracy in flux prediction.
}
\label{fig:ssi_index_model}
\end{figure}

The forecasting ability of the four models is evaluated in terms of the fitness between predicted flux and realistic flux, as shown in Fig.~\ref{fig:model_driving_time}, where the metric of distance is defined as the travel time (results with alternative distance metrics can be found in Fig. S2 and Fig. S3). As can be seen, independent of distance measures, extensions of the G-GM and dirG-GM models demonstrate better accuracy than GM. To further quantify the forecasting ability in four models with different distance metrics, an indicator named SSI based on the S\o rensen index~\cite{Lenormand_2012,liang2015a} is introduced to evaluate the model performance directly. In detail, the measurement index is defined as $SSI = \frac{2 \sum_{i} \sum_{j} \min(F_{i j}, \hat{F}_{i j})}{\sum_{i} \sum_{j} F_{i j} + \sum_{i} \sum_{j} \hat{F}_{i j}}$, where $F_{i j}$ and $\hat{F}_{i j}$ denote the actual flux and the forecast value of the size of the population migrating between city $i$ and city $j$, respectively. The range value for $SSI$ is 0, with a complete difference between the actual value and the predictive value of 1 with complete equality. A larger SSI means a better predictive capability of models and vice versa. From Fig.~\ref{fig:ssi_index_model}, it can be seen that the predictive performance of G-GM with travel time being a cost metric is the best approach among four models, and the SSI index of GM and aveG-GM is obviously worse. This finding further illustrates that GDP demonstrates adequate information about the city's ability to promote labour mobility, in contrast to per capita GDP, which is conventionally employed to measure demographic and economic influences simultaneously. Because of G-GM's stronger performance, the subjective consciousness of the workforce, e.g., pursuit of profit, is indeed profound in helping understand migration. Consistent with the previous study~\cite{Zhao_2016}, the appearance of travel time as the most appropriate distance metric also suggests that travel time weighs heavily in choosing migration destinations, instead of the geographical distance.

In addition, as reported in Table~\ref{tab:fit_parameter}, the exponents for the distances between cities and economic levels of origin and destination cities are all positive, suggesting that our assumption of more migration flows between close and well-developed cities may be justified. Additionally, the positiveness of $\gamma$ in all models clearly indicates that distance blocks labour mobility. However, aveG-GM has the worst performance, which suggests that GDP per capita is not a good choice. Note that $\alpha$ and $\beta$ are close to each other in dirG-GM (0.9278 and 0.9351 separately), implying that economic levels (GDP) of the origin and destination exert an approximately equivalent effect on the migrant population of the workforce, although the destination has a slightly stronger influence than the origin. Therefore, the direction of flows can reasonably be ignored, and dirG-GM can be reduced to G-GM. Therefore, combined with predictive performance and the simplicity of parameters, G-GM with travel time is the best solution to predict inter-urban migration.

\subsection{Mining inter-urban migration patterns}

Although the G-GM model takes economic status as a driving force in workforce migration, the sensitivity of individuals to travel costs and economic benefits is diversified because of different backgrounds. Thus, this sensitivity is significant for understanding workforce migration behaviours from a micro-perspective. We argue that instead of exploring this phenomenon at the collective level, focusing the analysis by mining workforce migration patterns of different sensitivities might suggest a new way to visualize the big picture of workforce migration.   
 
\begin{figure}[!h]
\centering
\includegraphics[width=7in]{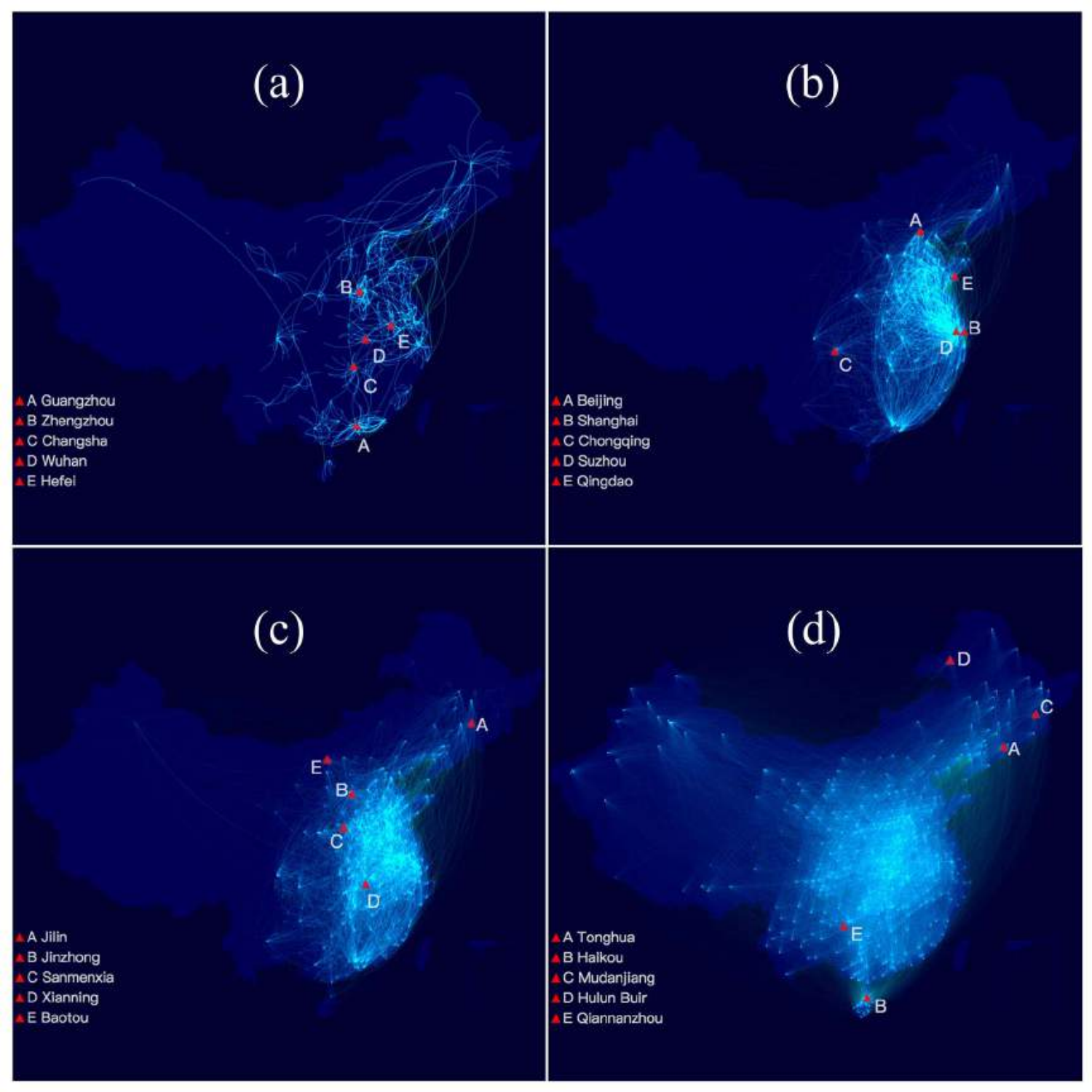}
\caption{
{\bf  Migration flows of grouped trajectories.}
The bright lines denote the migration flows between cities. (a) \textbf{Pattern I}, (b) \textbf{Pattern II}, (c) \textbf{Pattern III} and (d) \textbf{Pattern IV}. A brighter area denotes that there are more city pairs with migration flows. The average migration flows between city pairs is 18522, 3182, 1045 and 195, respectively. Each group accounts for 29\%, 28\%, 20\% and 23\% of the total flux, respectively. The representative cities in each pattern are marked by red triangles and are obtained by filtering the flux ratios of more than 50\% in \textbf{Patterns I \& II \& IV}(40\% in \textbf{Pattern III}), sorting each city's migration flux and selected cities.
}
\label{fig:group_trans_map}
\end{figure}

To effectively detect workforce migration patterns, an unsupervised approach of $K$-means is employed to cluster the migration trajectories of massive numbers of individuals (see Methods). As for identifying the optimal number of clusters, two methods, the silhouette coefficient~\cite{rousseeuw1987silhouettes} and elbow method~\cite{thorndike1953who}, are both used. We find that excellent clusterings can be achieved with three, four, or five clusters (see Fig. S4-S5). To obtain the best choice, different partitions are compared thoroughly, and the partition with four clusters demonstrates more patterns than that with three clusters and patterns with greater differences than that with five clusters (see Fig. S6-S7). Thus, the workforce is grouped into four explicable clusters, as shown in Fig.~\ref{fig:group_trans_map}.  

As can be seen, trajectories in four clusters have significantly different patterns from the perspectives of geography (see Fig.~\ref{fig:group_trans_map}), the product of GDP (see Fig. S8), travel time (see Fig. S9) and migration flux (see Fig. S10). All the patterns are depicted and explained as follows. 

\textbf{Pattern I Migration between local core cities and their surroundings}. As demonstrated in Fig.~\ref{fig:group_trans_map}(a), divergent, star-like patterns of trajectories indicate local yet core cities such as Guangzhou, Zhengzhou, Changsha, Wuhan, and Hefei, which are capital cities of provinces and which have a great economic influence within their neighbourhoods. The workforce in this group spends the least travel time and dominates the migration flux more than other groups. This finding implies that the essential component of migration flows between capital cities and their adjacent areas and travel time has an intensive effect on migration. The radiation to labour forces within relatively small areas of local core cities, especially capital cities, is reflected in this pattern. 

\textbf{Pattern II Migration between developed cities}. As can be seen in Fig.~\ref{fig:group_trans_map}(b), city pairs show the best economic statuses in this pattern, for instance, the most developed cities, such as Beijing, Shanghai, Chongqing, and Suzhou, are key destinations for the workforce of this group. This finding suggests that this pattern of workforce migration is more sensitive to cities' economic levels. Apart from that, the city pairs reflect that the workforce capacity in cities with high GDP match the requirements of the labour market of similarly developed cities and provide the workforce with more opportunities to obtain a lucrative job. As compared to \textbf{Pattern I}, migration flows in this pattern involve the second-lowest amount of travel time. Additionally, the mutual promotion of city pairs in this group, which are better developed, is fully reflected.   

\textbf{Pattern III Migration between undeveloped cities}. From the point of view of regional distribution, there are distinct differences between \textbf{Pattern III} and the former two patterns, and the bright region of this pattern concentrates mainly in central China, such as Jinzhong, Sanmenxia, and Xianning, among other cities. Cities experiencing this migration pattern are less extrusive than those in the former two groups, as can be seen in Fig. S8, and the product of GDP in these trajectories is relatively small. Based on the above discussion, we know that a city's GDP may not only have potential benefits for the labour market of this city but also represent technology capability requirements (see Fig.~\ref{fig:gdp_relation}). Therefore, the workforce in this pattern, lacking adequate technical skills, cannot meet the capacity requirements of other labour markets in cities with high GDP, as discussed in \textbf{Pattern II}, and must select other cities with lower GDP as the migration destination. In comparison to \textbf{Pattern II}, the filtering effect of markets is highlighted, and the intrinsic motivation for benefit appeals is limited by objective constraints.        

\textbf{Pattern IV Migration due to emotions}. This group's trajectories geographically cover all areas of China, particularly peripheral provinces such as Xinjiang, Tibet, Heilongjiang, and Hainan and other remote districts located on the country's borders, including the cities of Tonghua, Haikou, Mudangjiang, Hulun Buir, and Qiannanzhou. City pairs in this pattern possess the lowest product of GDP but the longest travel time compared to other groups. Meanwhile, the average volume of migration between city pairs in this pattern is the smallest, and the total flux in this pattern accounts for a small proportion of all migration volume, implying that these few ``emotional" migrations may be not completely profit-driven. Although some city pairs reflect labour migration to distant cities to escape harsh living environments, such as migration from Tibet and Xinjiang, much migration is closely related to tourism and sightseeing and protection from winter; the key cities in this pattern, Mudanjiang and Hulun Buir, are located in the cold region of northeast China, and Haikou and Qiannanzhou have a warmer southern climate.   

The dominant occupation of the workforce in \textbf{Pattern I} indeed illustrates the agglomeration effects of core cities to their surroundings in local districts. More surprisingly, these surroundings are not less economically developed cities, as intuitively expected, but demonstrate relatively better economic statuses within their provinces (GDP product ranks second in four groups; see Fig. S8). This pattern can also be further justified by the similar values for $\alpha$ and $\beta$ in dirG-GM (see Table~\ref{tab:fit_parameter}), suggesting the identical roles of origins and destinations in economic attraction. In fact, as super-cities or mega-cities, most capital cities expand continuously, according to the development report of the Chinese floating population in 2016, and our results fully confirm this point. To some extent, this finding implies that the organization pattern of urban groups within provinces in China is gradually formed. In addition, considering city-pairs with good economic development and located a short distance apart in \textbf{Pattern II}, the changes in labour demand are particularly apparent. With the development of these prosperous regions, labour with knowledge and techniques is more needed than simple manual labour, explaining the fact that few people in the workforce would like to move between less affluent cities located a long distance apart, as reflected in \textbf{Patterns III \& IV}. It is increasingly difficult to gain employment positions in developed areas because of workers' lack of skills. Thus, a workforce that pursues profit maximization not only needs a developed destination that provides many job opportunities but also needs an origin featuring a satisfactory economic level.

We further use G-GM to model the trajectories of different groups. As shown in Fig. S11 and Fig. S12, the prediction performance of grouped trajectories are all better than that of all trajectories, especially for the first two groups, which occupy the largest proportion (see also Fig. S10). The significantly enhanced forecasting ability suggests the rationality of clustering trajectories, and the workforce possesses different migration preferences (see also Fig. S11 and Fig. S12). In fact, individuals' sensitivity to migration costs can be sufficiently reflected by the gaming between economic benefits and migration costs, and the variations in $\gamma$ across groups are decent proxies to probe this phenomenon. As listed in Table~\ref{tab:group_fit_parameter}, $\gamma$ varies for different patterns and is clearly distinguished from the parameter of all trajectories, further implying the existence of workforce sensitivity to migration costs. Specifically, the small $\gamma$ (0.1910) for the first group indicates that labour is more sensitive to economic attraction and cares little about the travel time. With the incremental increase in travel time between cities, the sensitivity to travel costs is significantly improved; for example, in the second group, $\gamma$ rises to 0.3868, suggesting the priority of moving to well-developed neighbouring cities. Similarly, as economic statuses worsen in \textbf{Pattern III}, $\gamma$ jumps to 0.4186, reflecting the greatest sensitivity to travel time. 

However, the $\gamma$ for the last group actually falls with the longest travel time and the worst economic status compared to that of the other groups. Under this circumstance, the migration followed by the trajectories in \textbf{Pattern IV} are not affected by the economic benefit and the travel time, e.g., emotional motivation instead of profit-seeking motivation. Considering the specificity of the Spring Festival travel rush as a long holiday in winter, we conjecture that migration for nonprofit reasons, such as tourism sightseeing, protection from winter or visits with relatives, is filtered out and isolated in this cluster. For instance, as is well known, an increasing number of people from northeast areas have selected Hainan province as a place to seek refuge from the cold and to celebrate the Spring Festival in recent years. From Fig.~\ref{fig:group_trans_map}, we also find the trajectories between cities in Hainan are indeed prominent in \textbf{Pattern IV} compared to other groups; to a certain extent, this pattern confirms our speculation about escaping cold winters. 

\begin{figure}[!h]
\centering
\includegraphics[width=5in]{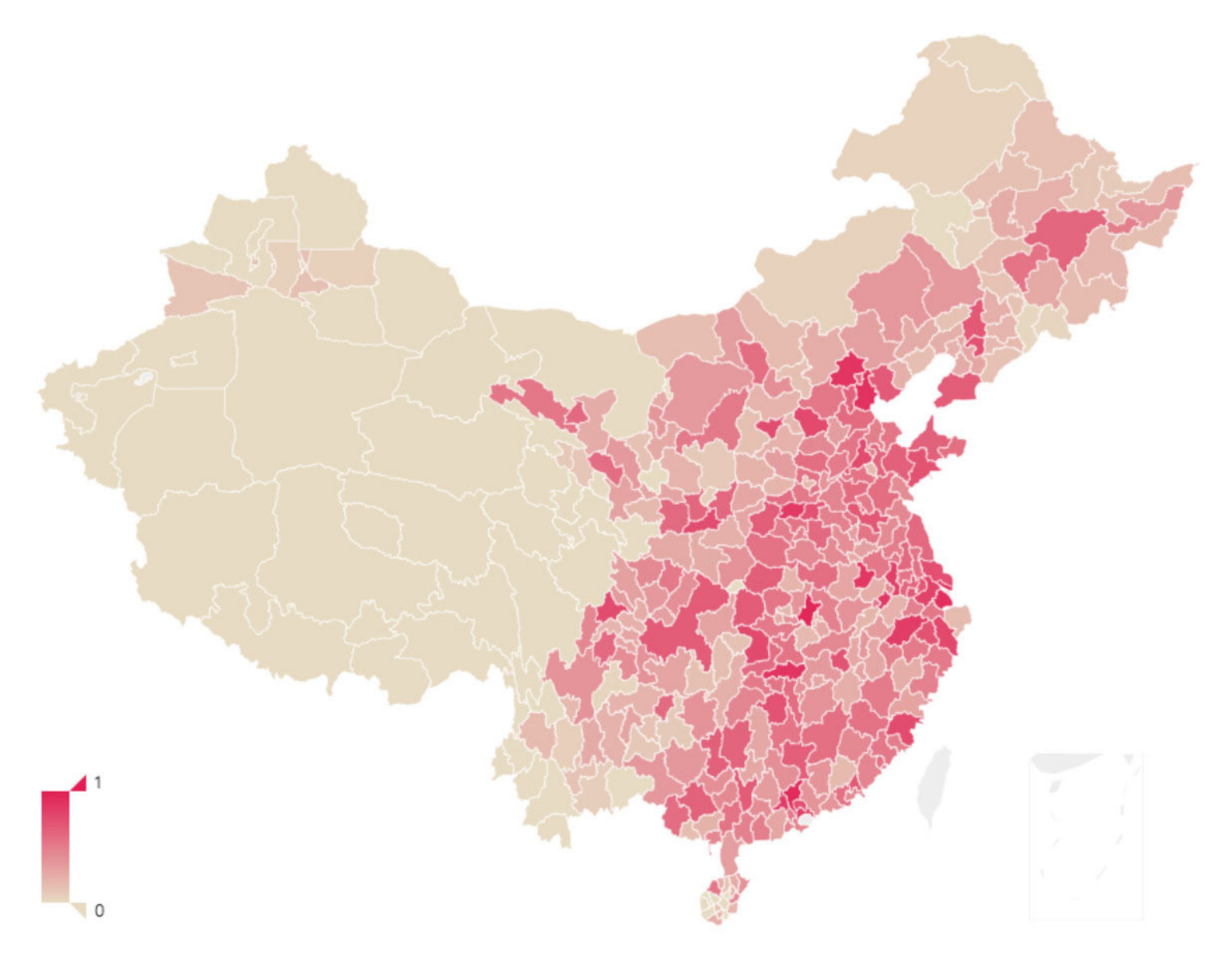}
\caption{
{\bf  Evaluation of cities' development models.}
The ratio flux in \textbf{Patterns I \& II} is defined as the index $DI$ to evaluate the development models of cities, $DI(i) = \frac{\sum_{(i,j) \in E(1) \cup E(2)} {F_{i j}}}{\sum_{(i,j) \in E}{F_{i j}}}$, where $E(1)$ is the set of city pairs in \textbf{Pattern I}, $E(2)$ is the set of city pairs in \textbf{Pattern II}, $E$ is the set of city pairs and $f_{i j}$ is the amount of workforce migration flux between city $i$ and $j$. A deeper red color indicates the city with a better development model. The value of $DI$ ranges from 0 to 1.
}
\label{fig:city_evaluate}
\end{figure}

\begin{figure}[!h]
\centering
\includegraphics[width=5in]{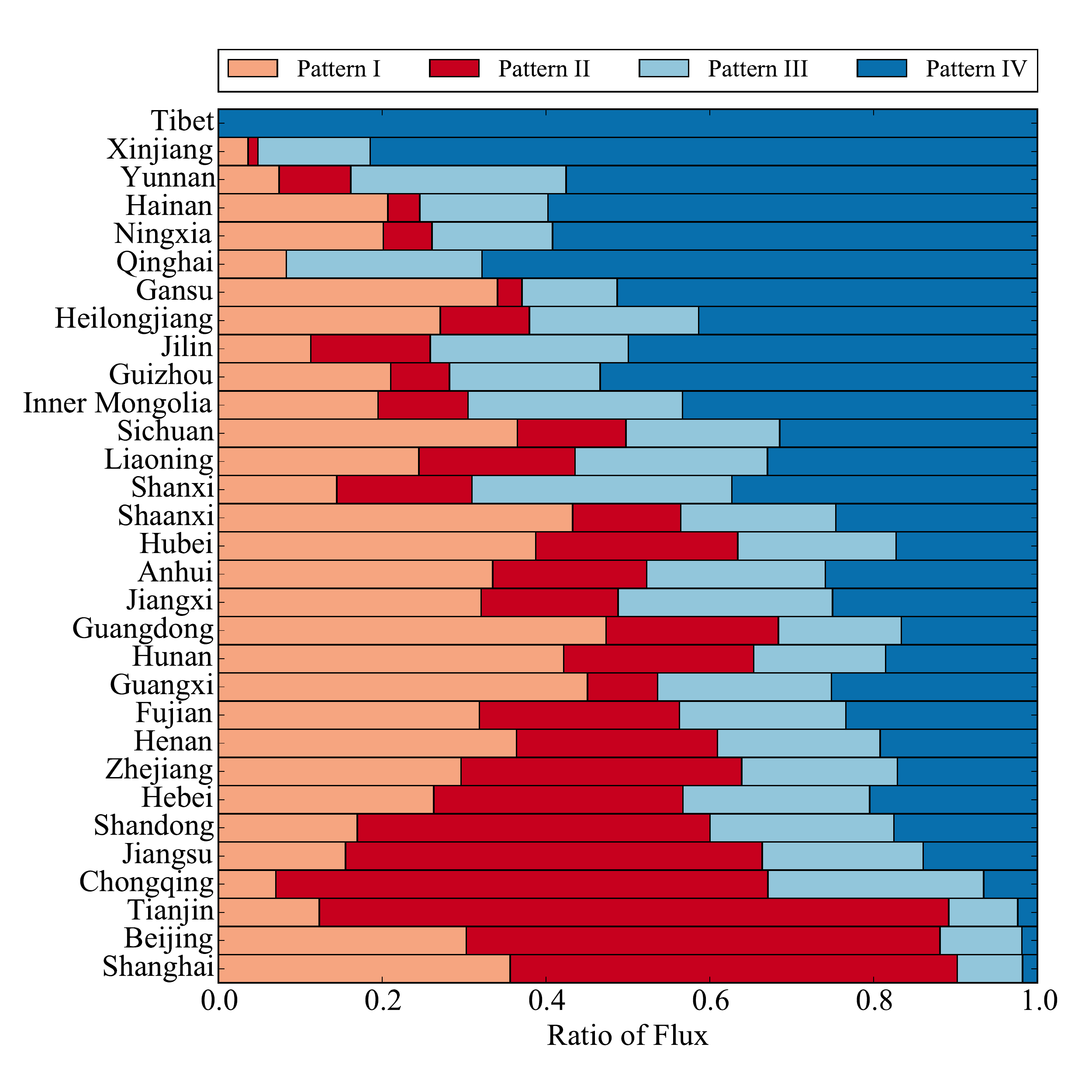}
\caption{
{\bf  Evaluation of provinces' development models.}
We define the average flux ratio in each pattern of cities in a province to evaluate the development model of the province. To be specific, $p(k, P)=\frac{\sum_{i \in P} \sum_{(i,j) \in T(k)}{F_{i j}}}{\sum_{i \in P} \sum_{(i,j) \in T} F_{i j}}$, in which $k$ stands for the pattern number, $P$ is the province and $i \in P$ means that city $i$ is under the jurisdiction of province $P$. $T$ represents all city pairs, and $T(k)$ is the set of city pairs belonging to Pattern $k$.
}
\label{fig:province_evaluate}
\end{figure}

In fact, \textbf{Pattern I} exactly embodies the development model for local core cities, e.g., their leading impacts on their surroundings. \textbf{Pattern II} further illustrates the mutual promotions among these similarly developed cities. In comparison to the other two patterns, the development models reflected in \textbf{Patterns I \& II} are more strongly expected to sustain growth. Therefore, the flux ratio of cities in \textbf{Patterns I \& II} reflects the sustainability of city development models, and we define this ratio as an indicator $DI$ to evaluate cities' development patterns. From Fig.~\ref{fig:city_evaluate}, we find that there are obvious variances in $DI$, and the cities with high development levels are scattered across the country. Additionally, from the view of the geographical distribution, the value of $DI$ in the country as a whole appears to significantly and progressively increase in a hierarchy from the western regions to the eastern areas, reflecting the disproportion and irrationality of development models in many regions, especially the western and northeastern areas. While unable to evaluate individual cities, this indicator is also applied to assess regional development models on a larger scale. Hence, the flux ratio in different patterns of provinces is used to further value the province development in Fig.~\ref{fig:province_evaluate}. There are significant differences among provinces in terms of the ratio of different patterns. The provinces with high ratios of \textbf{Patterns III \& IV}, as we well understand, are all undeveloped areas, such as Tibet, Xinjiang, Yunnan and etc.

\subsection{Regional development models inspired by migration patterns}
It is well known that urban agglomeration is the organization of space in a mature stage of city development. On one hand, because of their small-scale industries and relatively simple structures, small cities are not closely associated with greater production and operation networks and are considered to lack development vitality and momentum. There is an urgent need to optimize the allocation of a wider range of resources. On the other hand, from the perspective of big cities, many economic activities are no longer limited to a certain city with the growth of the urban population, the extension of industrial chains and the improvement of transportation infrastructures. Hence, it is necessary to strengthen ties among different cities in certain regions and create a collection of relatively win-win city pairs through appropriate regional development models. Meanwhile, urban diseases could be relieved by spatial expansion and the optimization of urban functions. Therefore, it is significant to promote regional development by making local cities play a leading role for neighbouring cities. More specifically, we hope for more \textbf{Pattern I} migration between city pairs.

\begin{figure}[!h]
\centering
\includegraphics[width=5in]{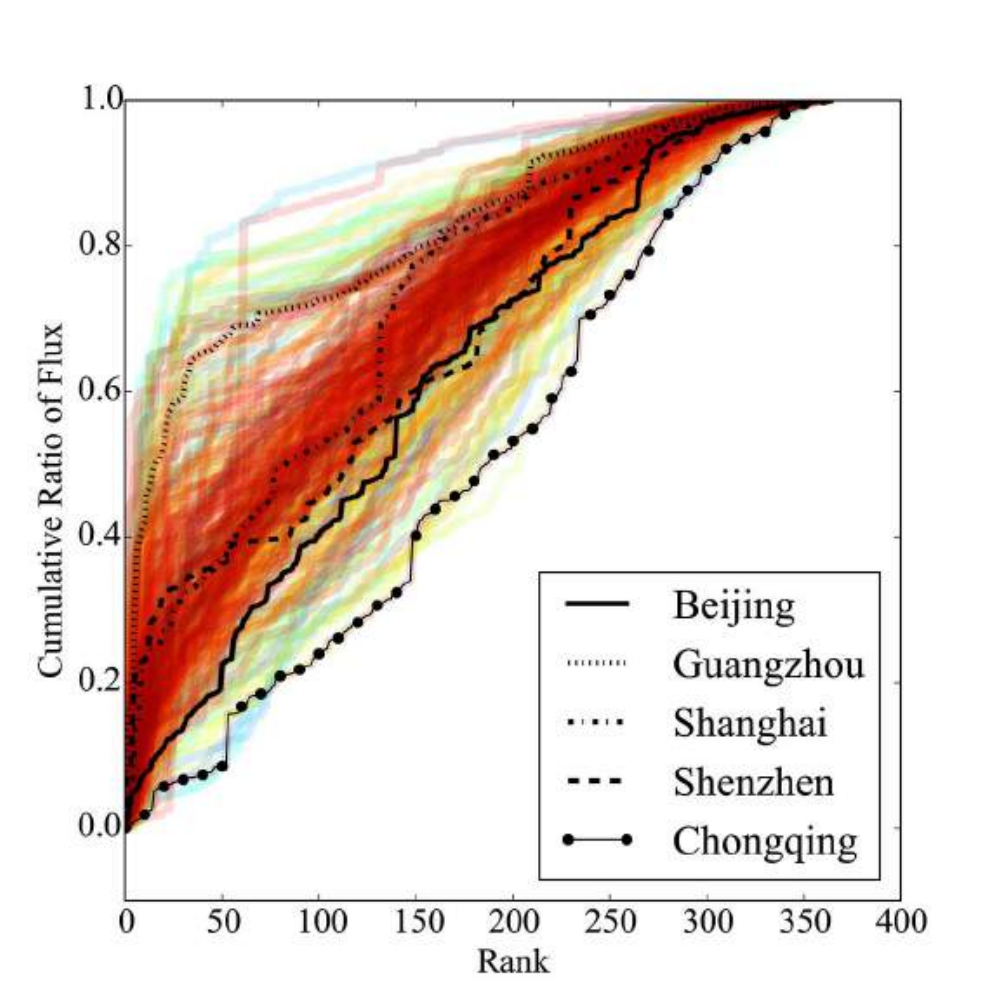}
\caption{
{\bf  Locality of cities.}
The cumulative ratio of the migration flux of the neighbouring cities from the near to the distant, defined as $c(i, r)$, is computed to show the locality characteristics of the target city. Specifically, $c(i, r) = \frac{\sum_{(i, j) \in T_i (r)}{F_{i j}}}{\sum_{(i,j) \in T_i}{F_{i j}}}$, where $T_i (r)$ is the set of city pairs between city $i$ and top $r$ cities, sorted by travel time. $T_i$ is the set of city pairs between city $i$ and other cities. The color of lines stands for the GDP of city $i$, and a deeper red color means better economic status.
}
\label{fig:city_ego_network}
\end{figure}

To understand the relation between cities and their surroundings in detail, we compute the cumulative ratio of flux from proximal to distant cities. This method is effective, universal and easy to execute in order to learn the present condition of the city development model. From Fig.~\ref{fig:city_ego_network}, we find that most cities are affected by their locality to some degree. In other words, for a certain city, more workforce members migrate from or to a closer city. Under these circumstances, local core cities could play a stronger role in radiating to peripheral cities, i.e., undeveloped cities could enhance their economic development by stimulating effects in nearby developed cities. At the same time, cities with unexpected development models are also detected; however, they lack locality and contribute little to regional development.

In terms of cities with good statuses, we find that most cities with high GDP have more workforce migration with neighbouring cities, as the red lines in Fig.~\ref{fig:city_ego_network} show. Nevertheless, unexpectedly, two developed cities, Beijing and Chongqing, are very different from other cities with high GDP because of their lack of locality. In particular, compared to Guangzhou, Shanghai and Shenzhen, the most important cities in China, Beijing and Chongqing, make an extremely low contribution to boost the development of their local regions. Our finding indicates that Hebei province, which is close to Beijing, is not similar to the Yangtze Delta Metropolitan and Pearl River Delta regions, whose rapid economic development is led by Shanghai and Guangzhou, respectively. Beijing, as the capital of China, has unique advantages in economic production resources and should play an significant lead role in radiating region development. As stated by a report on the government's work in 2014, achieving the coordinated development of Beijing, Tianjin and Hebei is a major national strategy. To some degree, our findings support the government strategy from the perspective of workforce migration. 

\begin{figure}[!h]
\centering
\includegraphics[width=5in]{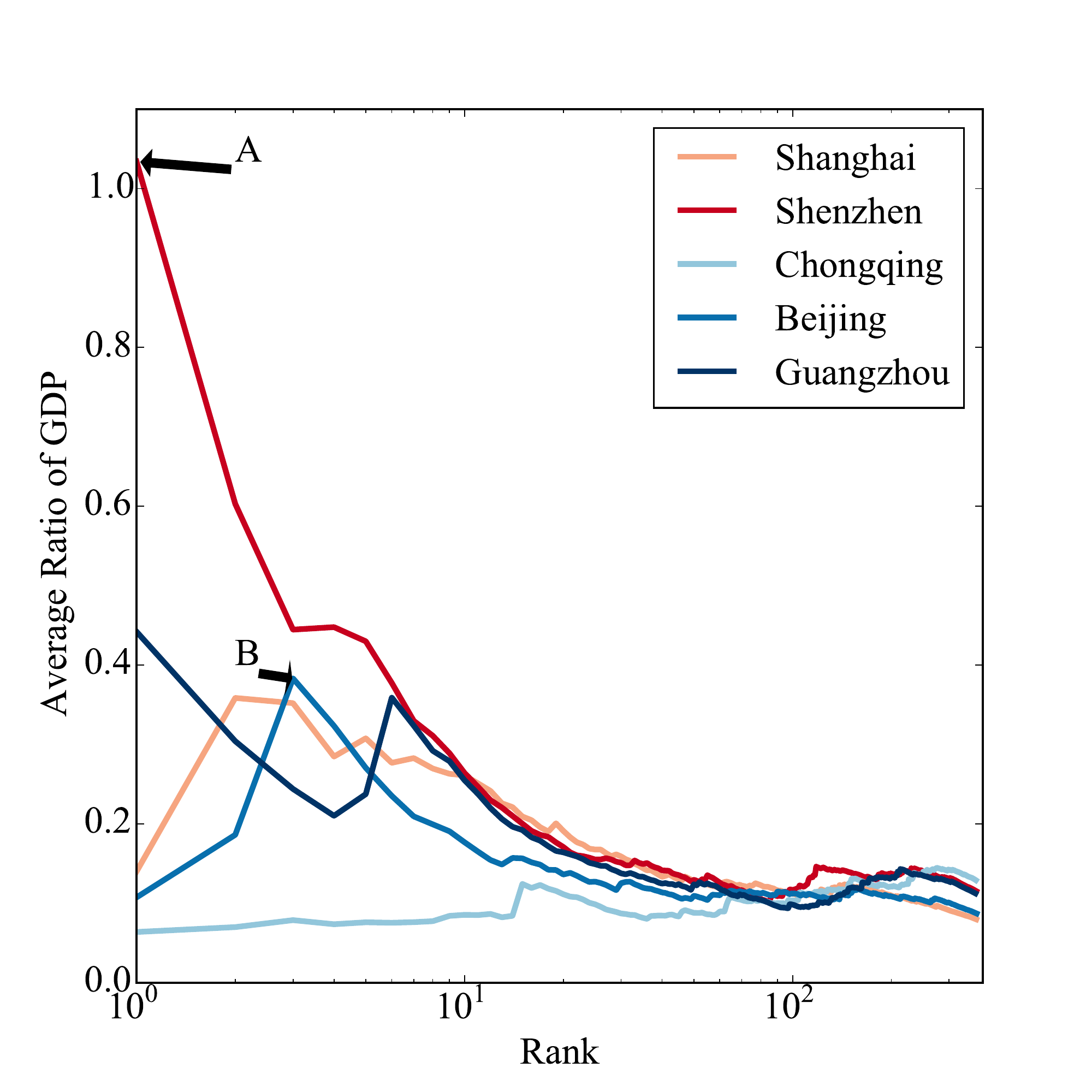}
\caption{
{\bf  Extent of GDP matching.}
To understanding the extent of GDP matching between target cities and their neighbouring cities, the average ratio of GDP is defined as $l(r, i) = \frac{\sum_{(i,j) \in T_i (r)}{E_j}}{r E_i}$, where $T_i (r)$ represents the set of pairs between city $i$ and top $r$ cities, sorted by travel time, and $E_i$ is the GDP of city $i$. Because Guangzhou has a higher GDP than Shenzhen, $A$ is an abnormal point in the line of Shenzhen. $B$ is the highest point in the line of Beijing, and the inflection point results from Tianjin, which is the only city with a high GDP that is similar to that of Beijing.
}
\label{fig:core_city_gdo_ratio}
\end{figure}

Furthermore, we analyze the influence factors for why Beijing lacks regional promotion relative to Guangzhou, Shanghai and Shenzhen. In Fig.~\ref{fig:core_city_gdo_ratio}, by comparing the GDP of core cities and their surroundings, we find that the matching extent of economic status between Beijing, Chongqing and its surroundings is worse than that of Shanghai, Shenzhen and Guangzhou. For Beijing, there is just one city, Tianjin, which has a matching economic status around the periphery. Additionally, the GDP of cities close to Chongqing are less than almost one-tenth of that city's GDP. Combined with the discussion above, due to the disparity between the technical level of the workforce in the cities' surroundings and the skill requirements of the labour markets in both Beijing and Chongqing, the labour mobility between these city pairs is discouraged, and hence, their radiation effects are essentially undermined. To promote regional development in a coordinated manner, enhancing the technical level of the workforce in undeveloped areas is a substantial requirement, except in policy-oriented economic cooperation.                                                  

\begin{figure}[!h]
\centering
\includegraphics[width=5in]{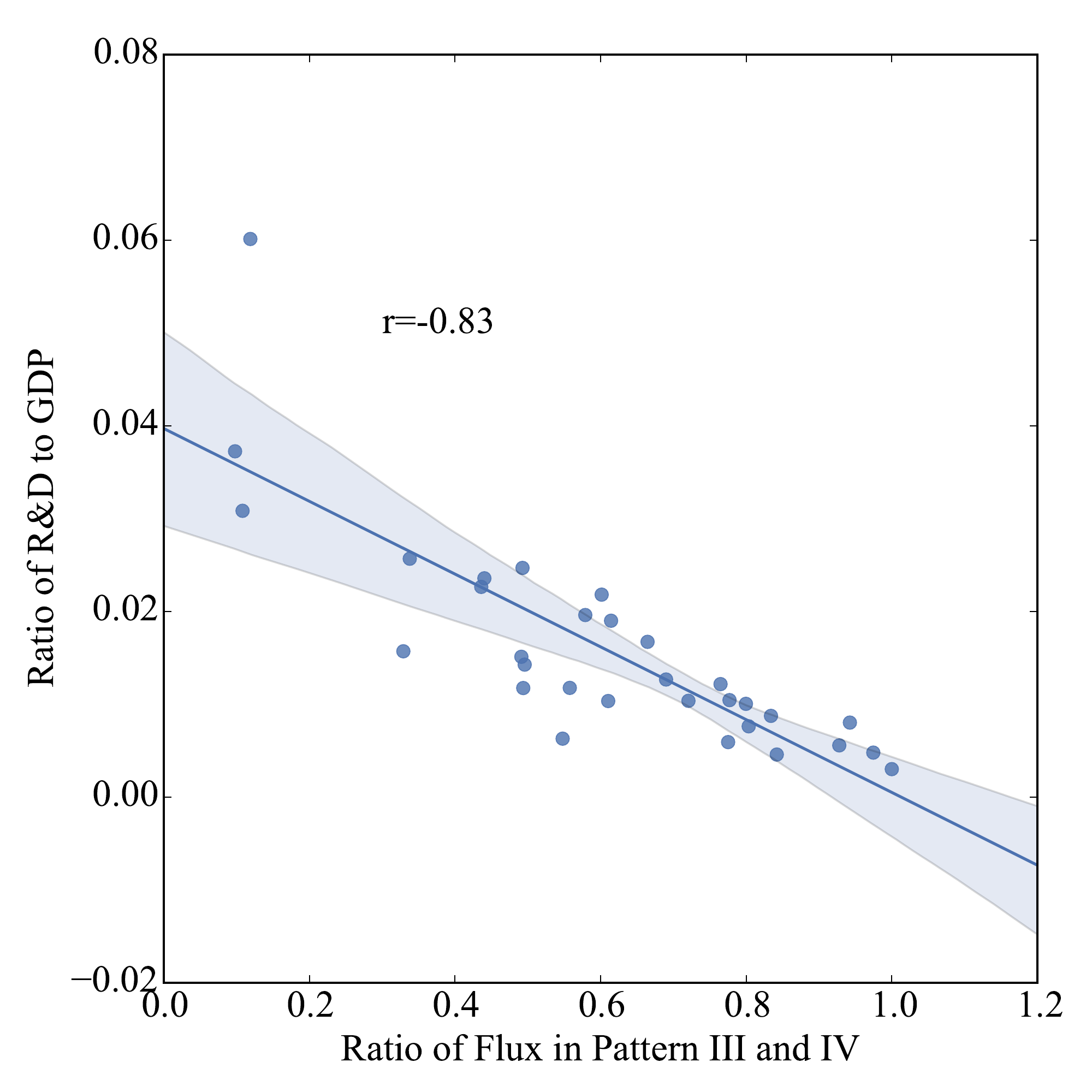}
\caption{
{\bf R\&D and the development models of cities.}
The ratio of flux in \textbf{Patterns III \& IV} is negatively correlated with the ratio of the investment of funds in R\&D to provinces' GDP.
}
\label{fig:rd_pattern_ratio}
\end{figure}

Almost all cities with a low GDP and a lack of locality, specifically, those ranking lower than Beijing, are in remote and undeveloped areas, for instance, Tibet, Xinjiang, Yunnan, and Guizhou. Combined with the discussion about \textbf{Patterns III \& IV} above, the workforce in these areas makes migration decisions under the threshold filtering in markets, migrating to other cities with similarly low GDP, or for ``emotional" reasons, neglecting the cost of the long-distance migration in order to escape their harsh life environments. To shift workforce migration from \textbf{Patterns III \& IV} to \textbf{Patterns I \& II}, a large, high-quality labour force is the key. From Fig.~\ref{fig:rd_pattern_ratio}, we observe that the ratio of the investment of funds in R\&D to GDP in provinces and the ratio of flux in \textbf{Patterns III \& IV} is negatively correlated (r=-0.83). This finding once again confirms the need for cities and provinces to vigorously expand scientific research and advanced education.  

\section{D\lowercase{iscussion}}

The social media boom indeed offers an opportunity to understanding human behaviours in unparalleled richness and fine granularity. To the best of our knowledge, this is the first study to systemically explore the workforce migration from the perspective of social media. Our study overcomes the limitations of traditional survey-based approaches and justifies the feasibility of probing workforce migration both collectively and individually using social media. More importantly, the present study confirms the possibility of quasi real-time policy making regarding social-economic issues, which are generally greatly constrained by the long census cycle and lead to the problem of being ``out-of-sync", which negatively influences the flexibility of policy formulation.          

Unlike previous models of human movement, in this study, we systematically discuss how the workforce migrates between cities within a country, considering the core driving force, interest articulation and the threshold filtering of the labour market. By introducing GDP as an economic indicator and travel time into the classical GM model, our G-GM model shows better performance than GM and justifies our assumptions. Additionally, due the diverse backgrounds of the workforce, the different migration patterns in terms of individual will and objective market requirements are mined and revealed. Meanwhile, we expound the gaming behaviour on benefits sought and travel cost evasion in exact patterns. Based on the patterns we find, a simple and effective method to evaluate city development models is proposed and provides support to find and solve realistic problems to enhance the sustainability of regional development. Our results also imply that human behaviors like collective movement can be essentailly entangled with economic development, and more inportantly, this entanglement can be sensed and prifiled through social media only.             

In addition to regional economic development, a demonstrated application discussed precisely in the paper, our study on workforce migration provides insightful ideas to solve a good deal of critical social-economic issues. The prediction model of migration flows offers a quantitative perspective to evaluate and design policies on household registration systems, floating population management and inter-city transportation planning in real time. Meanwhile, the disclosed patterns of workforce migration not only have implications for the government in policy formulation and evaluation but also inspire enterprises and individuals to solve real issues. For example, the gaming between the benefit and distance costs of making migration strategies could offer suggestions regarding where to release recruitment information to ease or solve the problems of labour shortages, especially in the southeast coastal areas of China. 

\section{C\lowercase{onclusion}}

In this study, geographical tweets posted on Weibo during the Spring Festival travel rush in 2017 are considered as a natural experiment on the national scale to understand the rules of workforce migration. Intrinsic motivation for the decision to migrate, such as the pursuit of interest and cost avoidance, are discussed systematically by introducing the economic indicators and travel costs into gravity laws. Additionally, a model called G-GM is presented to predict the migration flux. We find that the product of GDP and travel time between city pairs can be excellent indicators of workforce prediction; in particular, GDP reflects cities' potential benefits and the technical requirements of their labour markets. Under the combined effect of intrinsic motivation and external restrictions on the labour market, workforce migration in China presents four typical patterns. Moreover, individuals' sensitivity to migration costs can be well detected in terms of parameter fluctuation across different patterns. The migration flux ratio in divergent patterns provides us with a new bellwether for development models for cities or provinces. As an application, the practical problems and causes of regional development -- for instance, Beijing's and Chongqing's lack of a leading role in regional radiation and the mismatch between labour market skills in surrounding areas and labour market requirements in these two cities -- are detected from workforce migration patterns and inspire related policy formulation and evaluation. 

\section{M\lowercase{ethods}}

\subsection{Data description}
\emph{\textbf{Migration trajectories in Weibo}}. Tweets on Weibo during the Spring Festival travel rush, from January 13 to February 21, 2017, announced by the National Development and Reform Commission (see http://zizhan.mot.gov.cn/zfxxgk/)
, are thoroughly collected. Weibo is a Twitter-like service in China and its active users are 392 million, more than 330 million that of Twitter. Our dataset is a complete snapshot of Weibo during the travel rush. A 4-tuple $(m, u, t, g)$ is defined to represent a message with both temporal and spatial features, in which $m$ is the message number, $u$ is the unique identifier for an anonymous user, $t$ is the timestamp of the post and $g$ is the location, including the country, province and city information. Specifically, the posting city for each tweet can be further employed to establish the workforce migration network at the city level. Our dataset is publicly available and can be downloaded freely through https://doi.org/10.6084/m9.figshare.5513620.v2. 

\par\noindent\emph{\textbf{The national railway line data}}. We collect national railway line data on a total of 5,878 trains from a train schedule. For each city pair, we count the number of trains through two cities. The dataset is publicly available and can be downloaded through https://doi.org/10.6084/m9.figshare.5513620.v2.   

\par\noindent\emph{\textbf{Distance measures between cities}}. To measure inter-city accessibility, we use three metrics, the actual geographical distance, the travel distance and the travel time in route planning. The former is the distance computed by the longitude and latitude of cities on Google maps. More specifically, for city $i$, with longitude and latitude $(x_i, y_i)$, and city $j$, with $(x_j, y_j)$, the distance is defined as
\begin{eqnarray}
\label{geographical_dis}
    d_{i j} = D \times \arcsin( \sqrt {\sin ^2 (a)  + \cos(rx_i) \times \cos(ry_j) \times \sin ^2 (b)}  ),
\end{eqnarray}
 where $ry_i = \frac{y_i \times \pi} {180}$, $a = \frac {y_i - y_j}{2}$ and $rx_i = \frac{x_i \times \pi}{180}$, $b = \frac {x_i - x_j}{2}$, $D$ is the diameter of the earth. The travel distance and travel time between cities are crawled from the travel path planning in Baidu Map API. All distance measures between different city pairs are also publicly available and can be downloaded freely through https://doi.org/10.6084/m9.figshare.5513620.v2. 

\par\noindent\emph{\textbf{Demographic and economic surveys}}. To introduce the economic driving force of labour migration in our model, we collect the GDP and per capita GDP of all cities from a statistical yearbook of all provinces in China. In order to compare our model with the classical GM model, the population size in cities is also collected from a statistical yearbook of all provinces in China. Because of the restrictions in our household registration system, there is a significant difference between the population size by households registered and that by permanent residents. In our study, we choose the number of permanent residents to denote cities' population size. Because of the delay in publishing government statistics and the future forecasts using past data, we collect GDP, per capita GDP and the number of permanent residents in China in 2015 from the 2016 statistical yearbook. Furthermore, the ratio of practitioners in high-technology industries, the disposable income per capita and the investment of funds for research and development (R\&D) are all collected from the 2016 statistical yearbook. The datasets are publicly available and can be downloaded through https://doi.org/10.6084/m9.figshare.5513620.v2.

\subsubsection{Estimation of model parameters}

To obtain the parameters in the proposed models, we use the least squares polynomial fit. As an example, the process of solving the model dirG-GM $f_{ij} = a\frac{G_i^\alpha \cdot G_j^\beta}{d_{ij}^\gamma}$ will be discussed in more detail as follows. Because the flux between cities is greater than zero, we take its log and convert the model to a polynomial, as $\log f_{ij} = \log a + \alpha \log G_i + \beta \log G_j - \gamma \log d_{ij}$. To fit the parameters, the solution minimizes the squared error $E = \sum_{i,j} |p(i,j) - y_{ij}|^2$, where $p(i,j) = \tilde{\alpha} \log G_i + \tilde{\beta} \log G_j - \tilde{\gamma} \log d_{ij} + \log \tilde{a}$ and $y_{ij} = \log f_{ij}$.  

\subsubsection{Clustering migration trajectories}

Considering the diversity of human behaviour, we categorize workforce mobility into different kinds of movements by the clustering algorithm of $K$-means~\cite{MacQueen_1967,Anderberg_1973,Jain_1988}. In workforce trajectory set $D = \{\mathbf{p_1}, \mathbf{p_2}, \ldots, \mathbf{p_m}\}$, the trajectory between city $i$ and city $j$ is defined as a 6-tuple $\mathbf{p_k} = ( i, j, G_i, G_j, T_{i j}, F_{i j})$, where $G_i$ and $G_j$ denote the GDP of city $i$ and city $j$, respectively; $T_{i j}$ is the travel time in driving route plans between two locations, and $F_{i j}$ represents the population flow between city $i$ and $j$. The feature vector of trajectory $\mathbf{p}_k$ is defined as $\mathbf{v}_k = (\tilde{G}_{i j}, \tilde{T}_{i j}, \tilde{F}_{i j})$, where $\tilde{G}_{i j}$ is the normalization of the product of $G_i$ and $G_j$, $\tilde{G}_{i j } = \frac{G_i \cdot G_j - \min(G \cdot G)}{\max(G \cdot G) - \min(G \cdot G)}$; $\tilde{T}_{i j} = \frac{T_{i j} - \min(T)}{\max(T) - \min(T)}$ and $\tilde{F_{i j}} = \frac{F_{i j} - \min(F)}{\max(F) - \min(F)}$. The labour trajectory set can be divided into $n$ clusters $\emph{C} = \{C_1, C_2, \ldots, C_n\}$ by minimizing $E = \sum_{k=1}^{n} \sum_{\mathbf{v} \in C_k} \emph{dist}(\mathbf{v}, \mathbf{m}_k)^2$, where $C_k$ is the $k$th cluster, $\mathbf{m}_k$ denotes the center of cluster $C_k$. In detail, $\mathbf{m}_k = \frac{1}{|C_k|} \sum_{\mathbf{v} \in C_k} \mathbf{v}$. In this paper, we employ cosine similarity to measure the distance between two workforce trajectories, $\emph{dist}(\mathbf{v}, \mathbf{m}_k) = \frac{\mathbf{v} \cdot \mathbf{m}_k}{\parallel\mathbf{v}\parallel \parallel\mathbf{m}_k\parallel}$, where 
$||\mathbf{v}|| = \sqrt{\tilde{G}_{i j}^2 + \tilde{T}_{i j}^2+\tilde{F}_{i j}^2}$, and the optimization function is converted to maximize $E=\sum_{k=1}^{n} \sqrt {\sum_{\mathbf{v, u} \in C_k} sim(\mathbf{v, u})}$. The CLUTO (http://glaros.dtc.umn.edu/gkhome/views/cluto) package is employed in our study to conduct the clustering.

\section*{C\lowercase{ompeting interests}}
We have no competing interests.

\section*{A\lowercase{uthor contributions}}
H.X. and W.J. conceived the research. H.X. and Z.J. conducted the experiments and analysed the results. H.X. and Z.J. wrote the manuscript. All authors reviewed the manuscript.

\section*{A\lowercase{cknowledgments}}
Dr.Junjie Wu was supported by the National Natural Science Foundation of China (NSFC) (71531001, 71725002,
U1636210, 71471009, 71490723) and the Fundamental Research Funds for Central Universities. Z.J. thanks the support from NSFC(71501005).

\clearpage


\clearpage
\begin{table}
\centering
\begin{tabular}{| c | c | c | c | c | c |}
\hline
Distance metric& Model & $\gamma$ & $\alpha$ & $\beta$ & Prob(F-statistic)\\
\hline
\multirow{4}{*}{Geographical distance} & GM & $0.3137^{***}$ & - & -   & $^{**}$ \\
\cline{2-6}
& aveG-GM & $0.7574^{***}$ & - & -  & $^{**}$ \\
\cline{2-6}
& G-GM & $0.2872^{***}$ & - & - & $^{**}$ \\
\cline{2-6}
& dirG-GM & $0.3066^{***}$ & $\bf{0.9278}^{***}$ & $\bf{0.9351}^{***}$ & $^{**}$ \\
\hline
\multirow{4}{*}{Travel distance} & GM & $0.3910^{***}$ & - & -   & $^{**}$\\
\cline{2-6}
& aveG-GM & $0.8608^{***}$ & - & -  & $^{**}$\\
\cline{2-6}
& G-GM & $0.3409^{***}$ & - & - & $^{**}$ \\
\cline{2-6}
& dirG-GM & $0.3373^{***}$ & $\bf{0.9018}^{***}$ & $\bf{0.9088}^{***}$ & $^{**}$ \\
\hline
\multirow{4}{*}{\bf{Travel time}} & GM & $0.3901^{***}$ & - & -  & $^{**}$ \\
\cline{2-6}
& aveG-GM & $0.7307^{***}$ & - & - & $^{**}$ \\
\cline{2-6}
& G-GM & $0.3165^{***}$ & - & - & $^{**}$ \\
\cline{2-6}
& dirG-GM & $0.3628^{***}$ & $\bf{0.9167}^{***}$ & $\bf{0.9239}^{***}$ & $^{**}$ \\
\hline
\end{tabular}
\caption{
\label{tab:fit_parameter}The fitting results for different models. In this table, $\gamma$ is the exponent for the distance measurements, and $\alpha$ and $\beta$ are exponents for the GDP of origins and destinations, respectively. Additionally, $^{**}$ stands for $p<0.01$, and $^{***}$ represents $p<0.001$, all parameters in the models have been examined by significance tests.
}
\end{table}

\begin{table}
\centering
\begin{tabular}{| c | c | c | c | c | c |}
\hline
Trajectory set & All & Group I & Group II & Group III & Group IV \\
\hline
$\gamma$ &   $0.3165^{***}$ & $0.1910^{**}$ & $0.3868^{***}$ & $0.4186^{***}$ & $0.1961^{***}$\\
\hline
Prob(F-statistic) & $^{***}$ & $^{**}$ & $^{***}$ & $^{***}$ & $^{***}$\\
\hline
\end{tabular}
\caption{
\label{tab:group_fit_parameter} $\gamma$ varies in different migration patterns and all trajectories. $^{**}$ stands for $p<0.01$, and $^{***}$ represents $p<0.001$.
}
\end{table}

\clearpage

\clearpage
\newcommand{\beginsupplement}{%
                   \setcounter{table}{0}
                   \renewcommand{\thetable}{S\arabic{table}}%
                   \setcounter{figure}{0}
                   \renewcommand{\thefigure}{S\arabic{figure}}%
}

\beginsupplement

\begin{figure}[!h]
\includegraphics[width=7in]{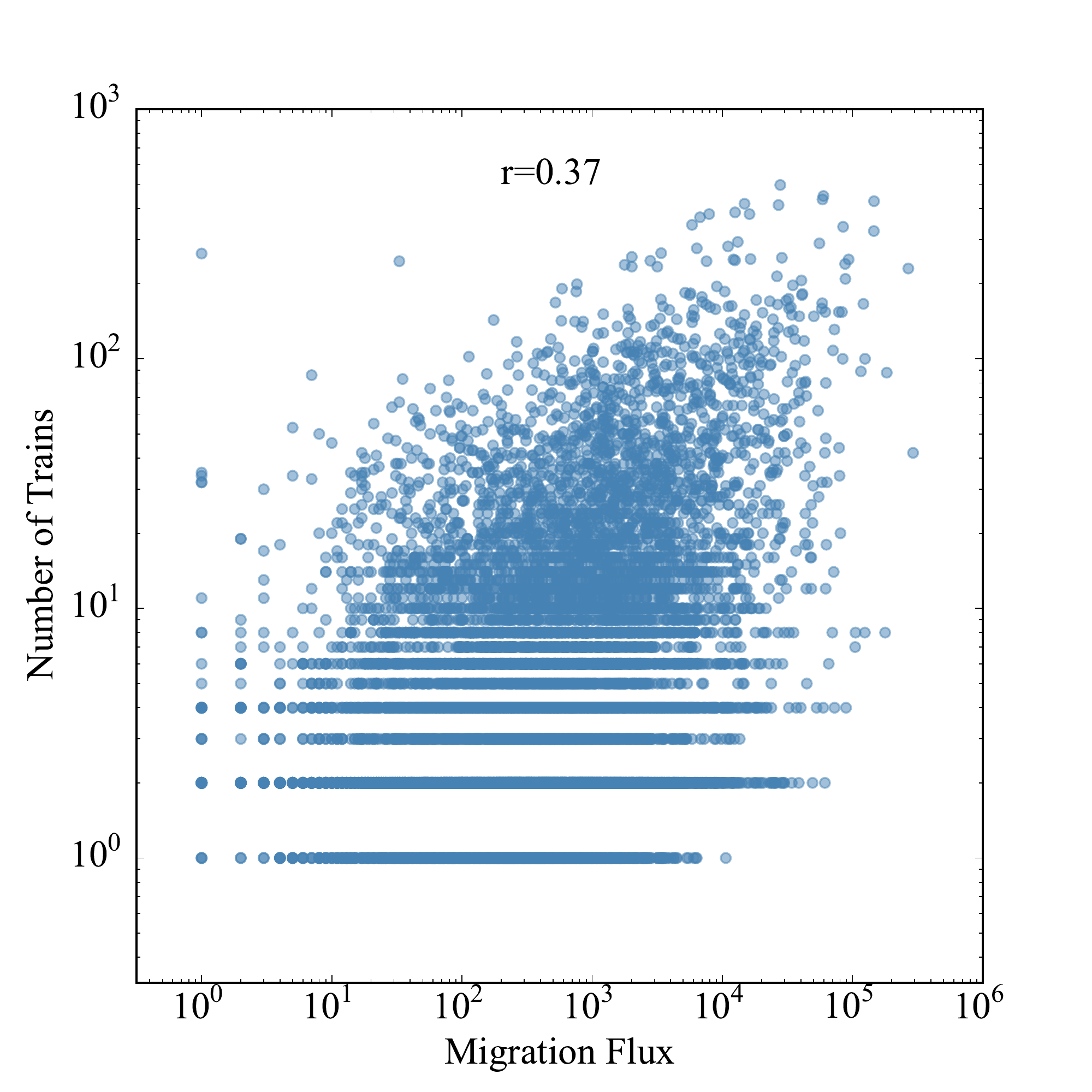}
\caption{
{\bf Relation between migration flux and the number of trains between city pairs.}
The migration flux is positively correlated with the number of trains that pass through city pairs, and the coefficient of correlation is approximately 0.37.
}
\label{fig:relation_migration_train}
\end{figure}

\begin{figure}[!h]
\includegraphics[width=7in]{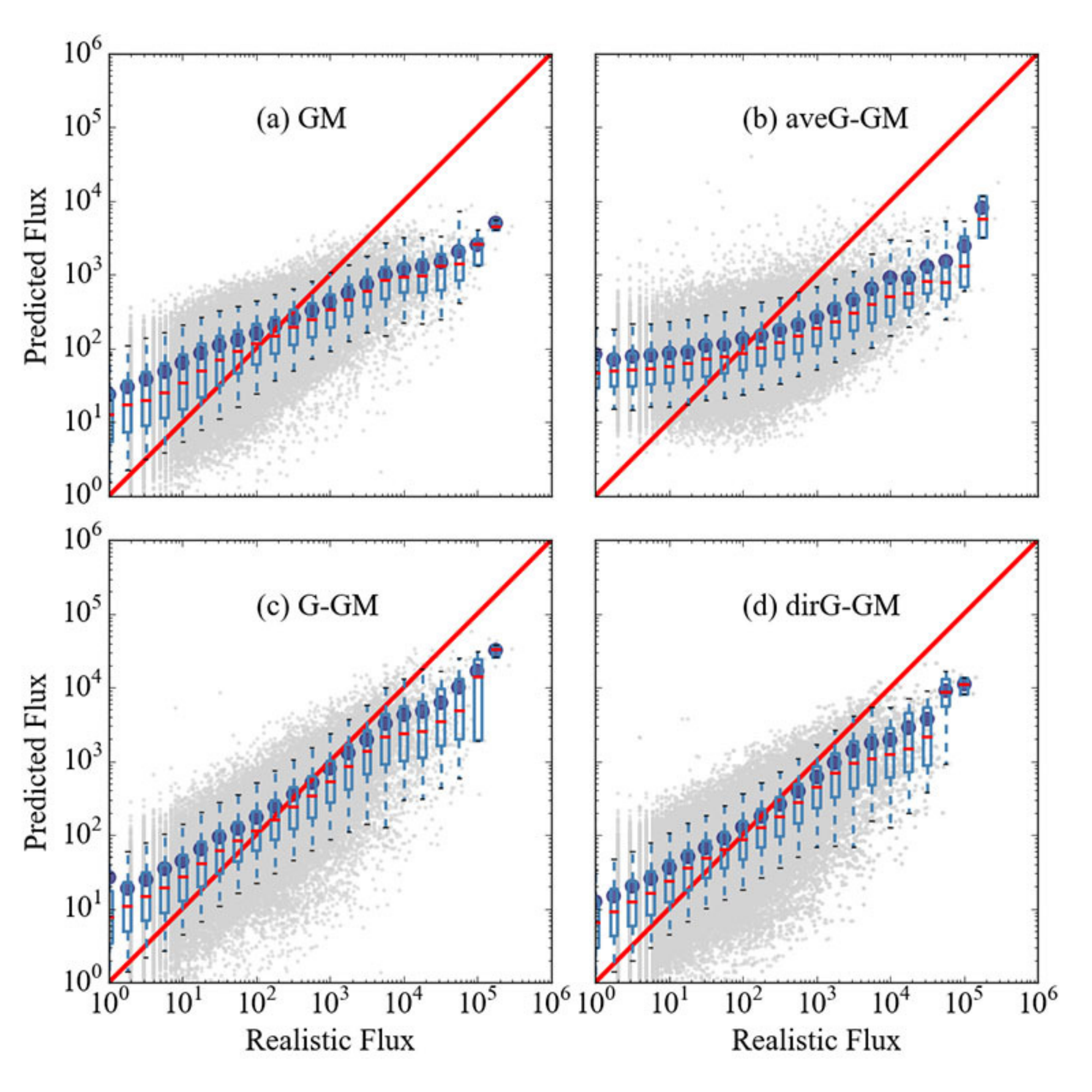}
\caption{
{\bf Modelling migration flows with the distance metric of geographical distance.}
The comparison between actual and predicted labour migration flows in four models.
}
\label{fig:model_geo_dis}
\end{figure}

\clearpage

\begin{figure}[!h]
\includegraphics[width=7in]{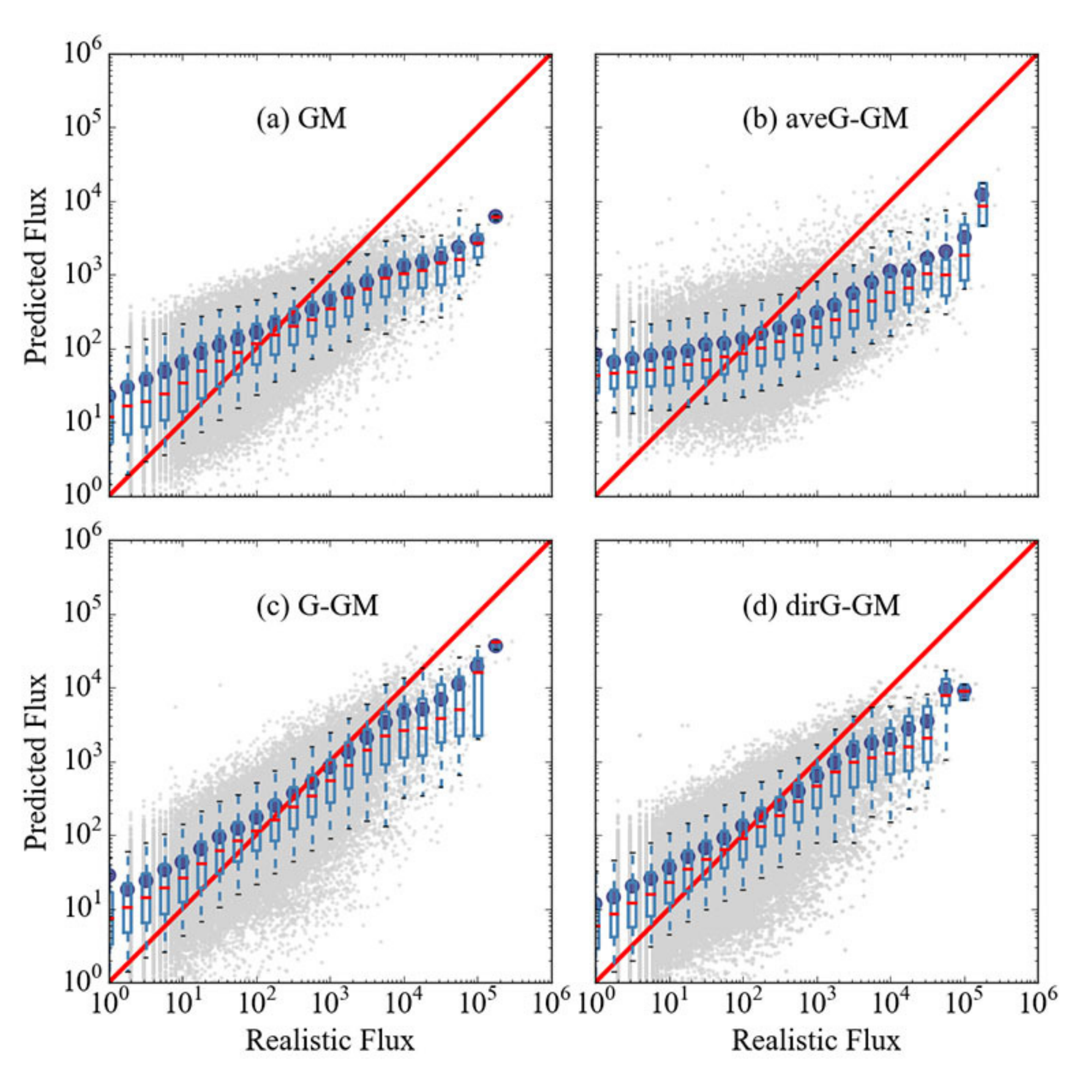}
\caption{
{\bf Modelling migration flows with the distance metric of travel distance.}
The comparison between actual and predicted migration flows in four models.
}
\label{fig:model_driving_dis}
\end{figure}

\clearpage

\begin{figure}[h!]
\centering
\includegraphics[width=6in]{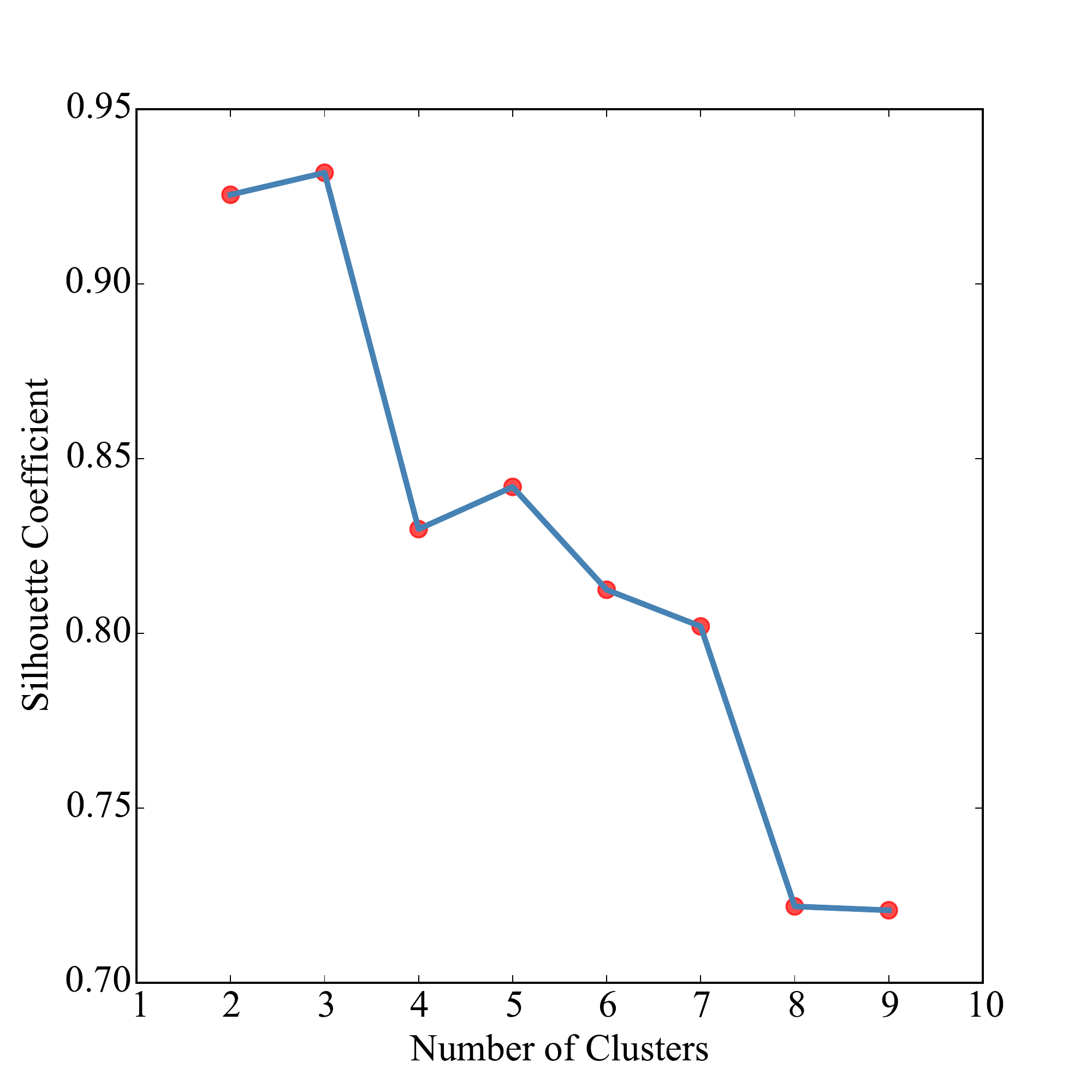}
\caption{
{\bf Comparison of different cluster numbers based on the silhouette coefficient.}
The silhouette coefficient for each sample $i$ is calculated as $s(i) = \frac{b(i) - a(i)}{\max(a(i) , b(i))}$, where $a(i)$ reflects the average dissimilarity between $i$ and all other samples within the same cluster and $b(i)$ is the lowest average dissimilarity between $i$ and any other cluster. In this paper, the cosine distance is defined as 1 minus the cosine similarity. Additionally, the average $s(i)$ over all samples is a metric to measure how appropriately the data have been clustered, with a value ranging from -1 to 1. A higher value denotes a better clustering result. It can be seen that the best effect (silhouette coefficient is 0.93) is achieved when the number of clusters is 3.
}
\label{fig:si_score}
\end{figure}

\clearpage
\begin{figure}[h!]
\centering
\includegraphics[width=7in]{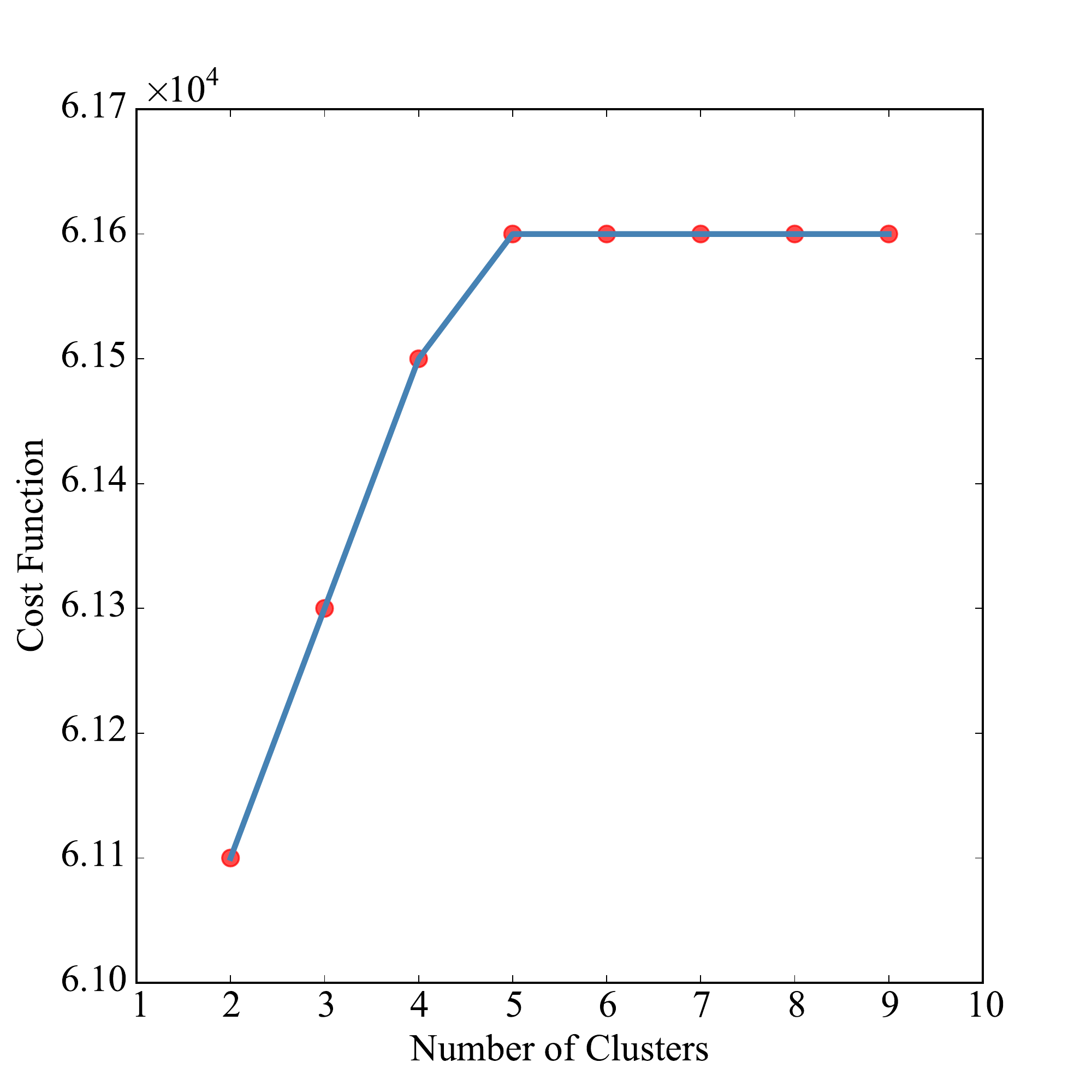}
\caption{
{\bf Comparison of different cluster numbers based on the elbow method.}
The elbow method is used to find the number of clusters when an obvious inflection point is obtained in the cost function. In this paper, the cost function maximizes the cosine similarity of samples within the same cluster. As can be seen, the best number of clusters is chosen at this point, where the number of clusters is set to 5.
}
\label{fig:kj_score}
\end{figure}

\clearpage
\begin{figure}[h!]
\centering
\includegraphics[width=7in]{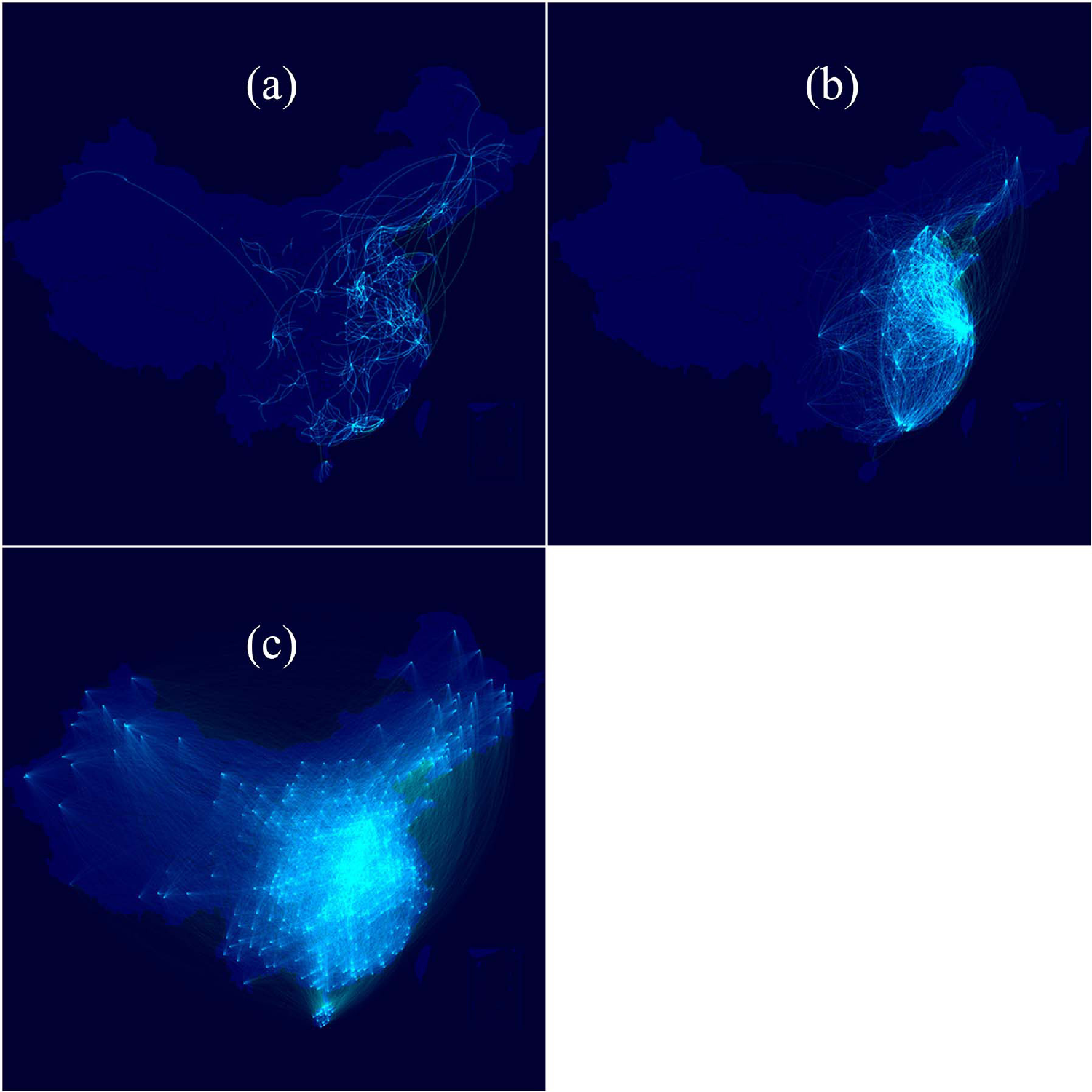}
\caption{
{\bf Migration flows of grouped trajectories with three clusters.}
(a) Pattern I; (b) Pattern II; and (c) Pattern III. Compared to the distinct patterns of four clusters, the partition with three clusters is not abundant enough. Although Pattern I and Pattern II can correspond to the first two patterns of the four-cluster case, Pattern III gives us an intangible result, including the brighter area in the midlands and the transition relationship between distant city pairs that are divided into two patterns in the partition of four clusters. Therefore, it is better to divide trajectories into four clusters because it offers more information.
}
\label{fig:map_3}
\end{figure}

\clearpage
\begin{figure}[h!]
\centering
\includegraphics[width=5in]{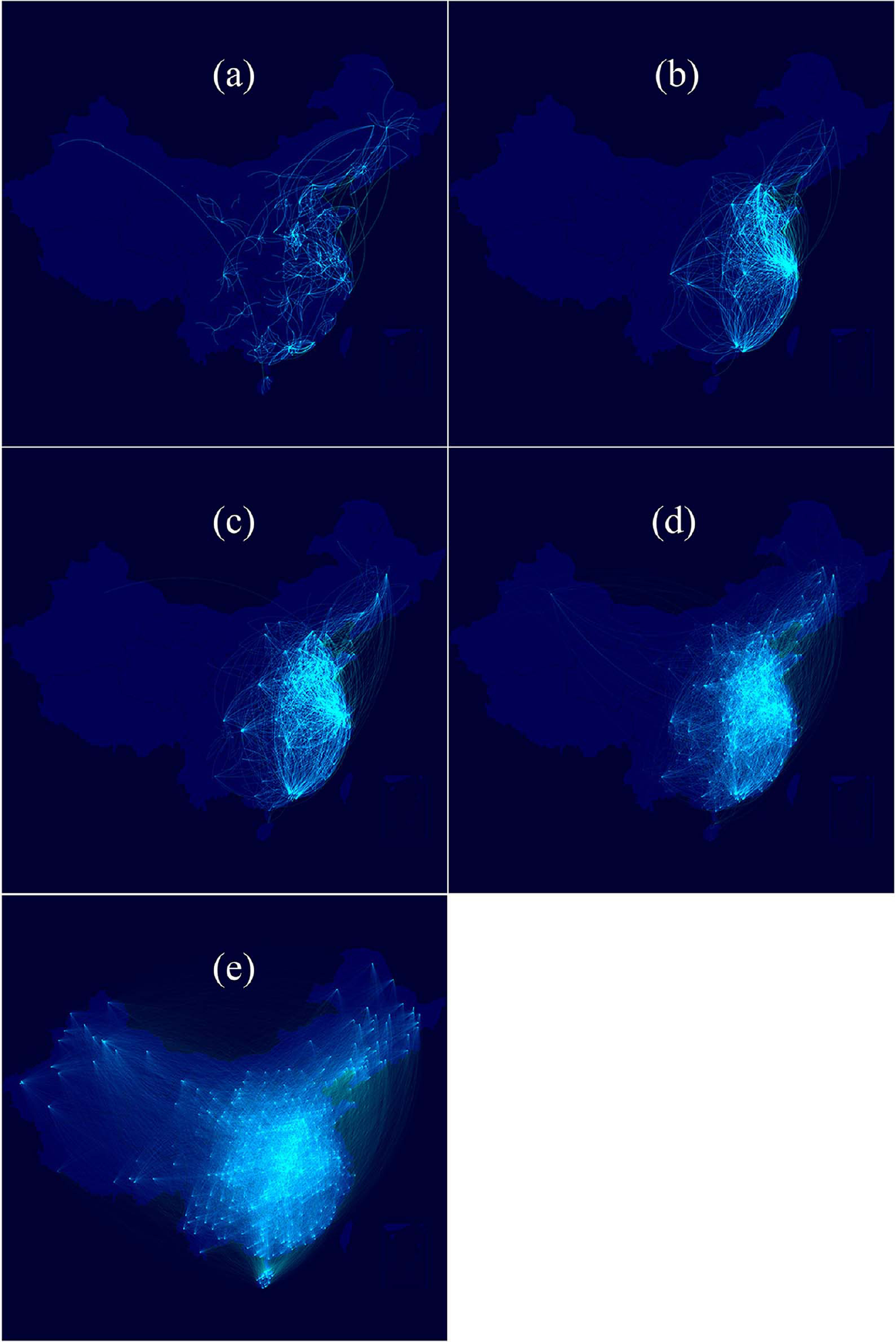}
\caption{
{\bf Migration flows of grouped trajectories with five clusters.}
(a) Pattern I; (b) Pattern II; (c) Pattern III; (d) Pattern IV; (e) Pattern V. There is no obvious distinction among Patterns II and III. Compared with this partition result, the patterns of the four clusters are obviously more different from each other. Accordingly, the partition with four clusters is a better choice than that with five clusters.
}
\label{fig:map_5}
\end{figure}

\clearpage
\begin{figure}[h!]
\centering
\includegraphics[width=7in]{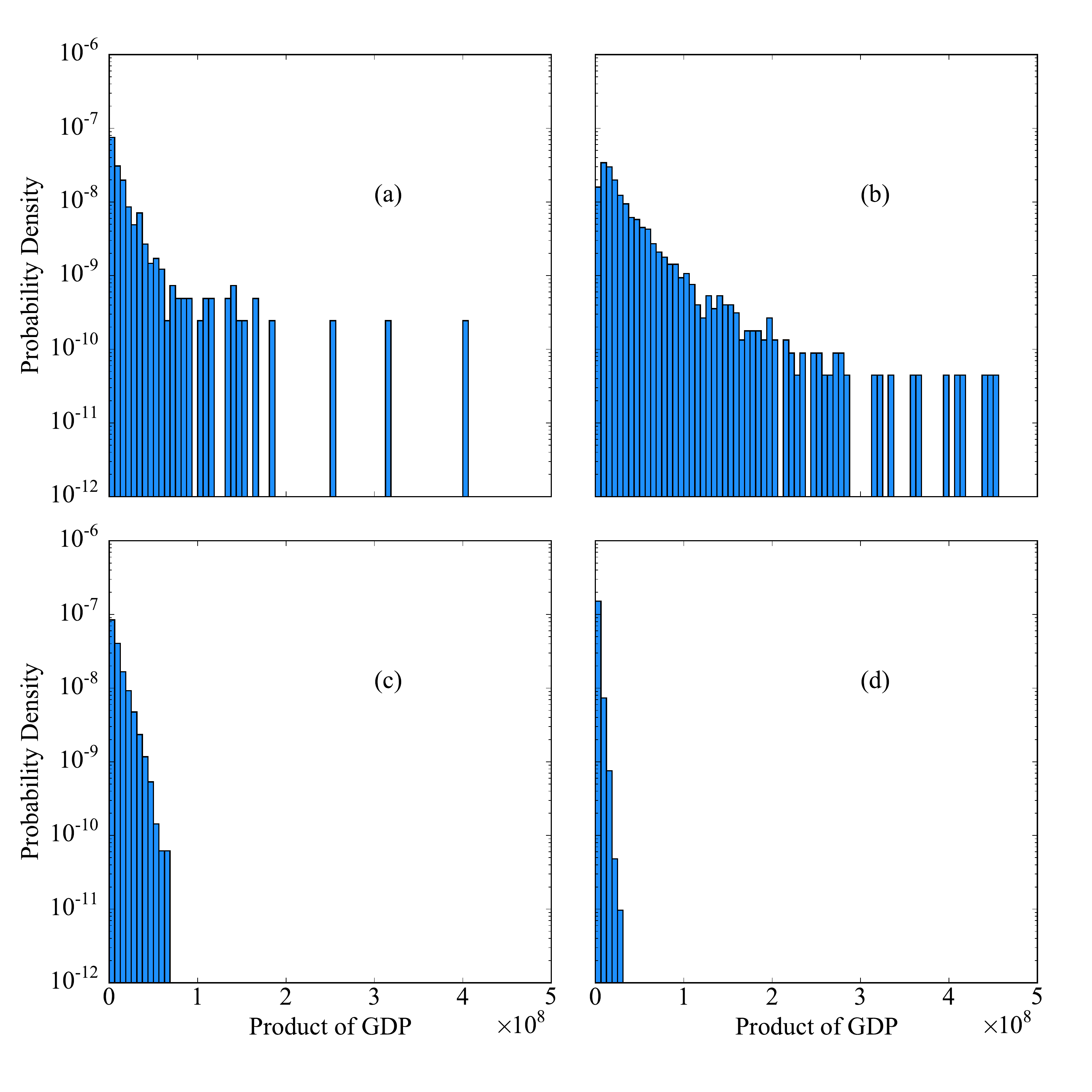}
\caption{
{\bf GDP product distributions in grouped trajectories.}
(a) Pattern I; (b) Pattern II; (3) Pattern III; (4) Pattern IV. The highest probability of migration occurs between two cities with excellent economic levels in the second cluster, and the first subgroup and the third subgroup follow, with the second- and third-highest probabilities. In comparison, almost all city pairs in the fourth group have low levels of economic development.
}
\label{fig:gdp_product_dis}
\end{figure}

\clearpage
\begin{figure}[h!]
\centering
\includegraphics[width=7in]{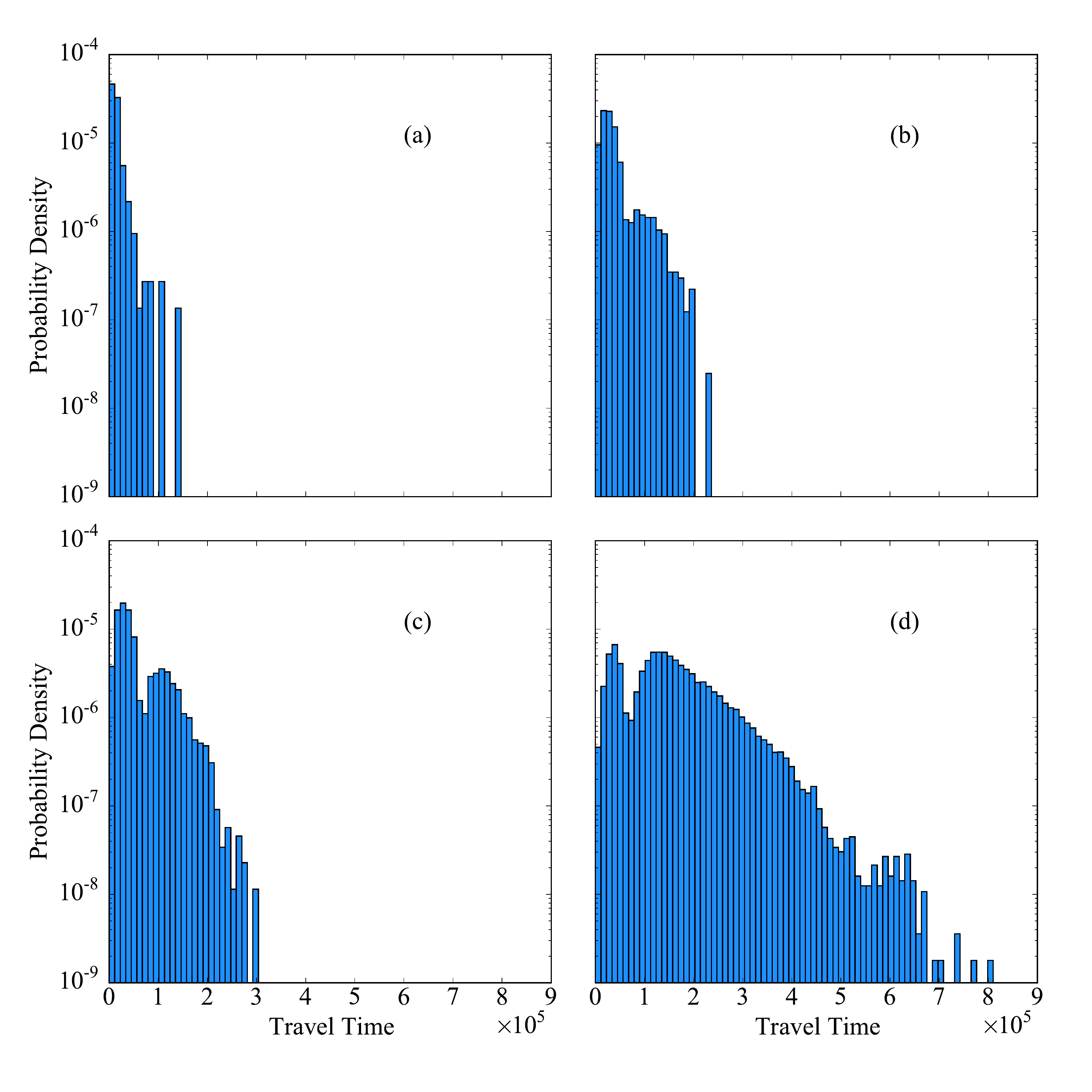}
\caption{
{\bf Distributions of travel time between cities in subgroups.}
(a) Pattern I; (b) Pattern II; (c) Pattern III; (d) Pattern IV. It is apparent that the city pairs in the first group have the shortest
travel time. Additionally, there is an increase from the first group to the fourth group, where travel time reaches the maximum.
}
\label{fig:drive_time_dis}
\end{figure}

\clearpage
\begin{figure}[h!]
\centering
\includegraphics[width=7in]{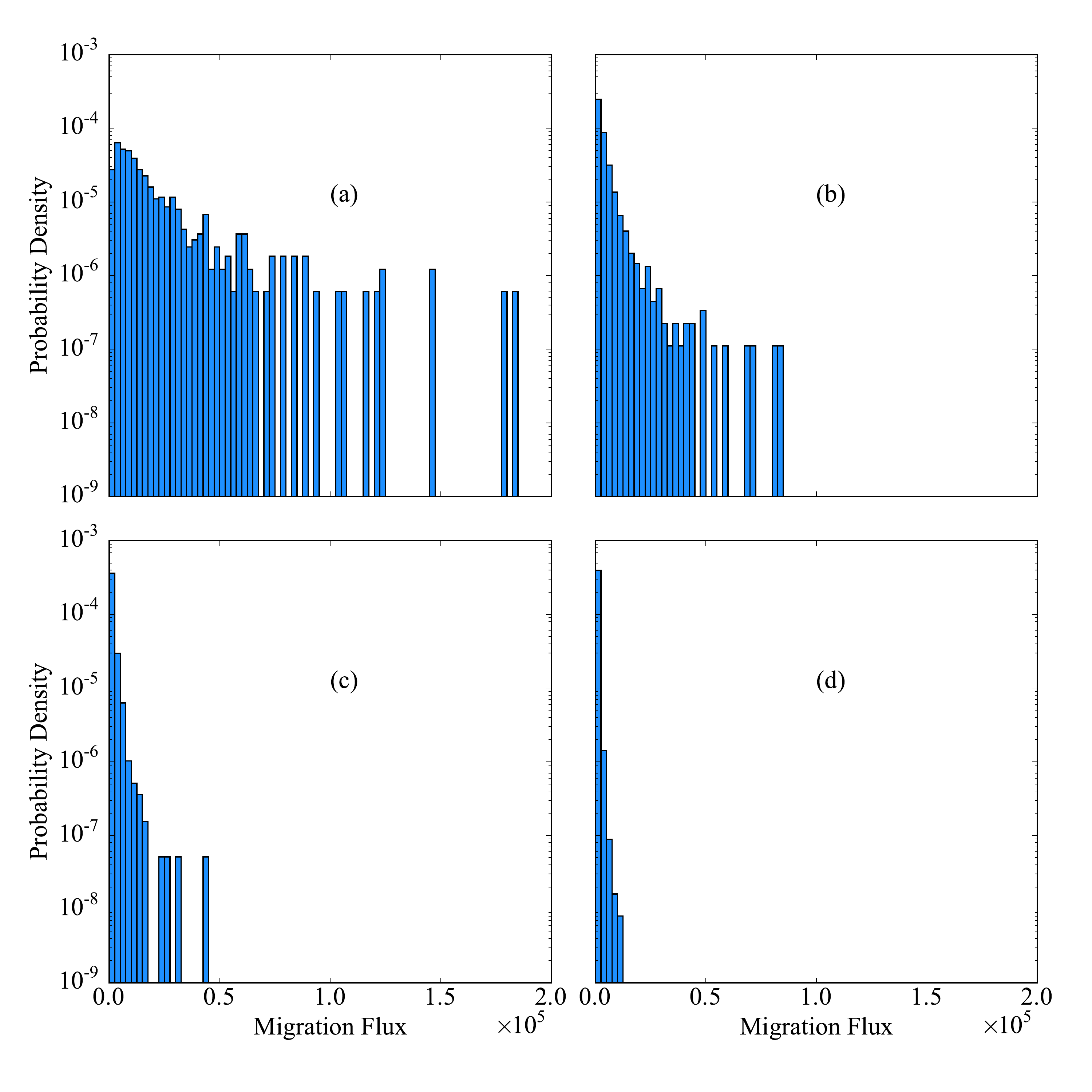}
\caption{
{\bf Distributions of migration flux between cities in grouped results.}
(a) Pattern I; (b) Pattern II; (c) Pattern III; (d) Pattern IV. There is a dramatic decline in the average migration flows between city pairs from the first group to the fourth group (18522, 3182, 1045 and 195 in detail). Additionally, the proportion of total flux in each group is 29\%, 28\%, 20\% and 23\% separately. Patterns I, II and III indeed explain the extremely weighty types of workforce migration behaviours. The scarce flow in each city pair and a small percentage of total flux in Pattern IV imply that the migration in this cluster can be regarded as noisy movements to some extent.
}
\label{fig:flux_dis}
\end{figure}

\clearpage
\begin{figure}[!h]
\centering
\includegraphics[width=6in]{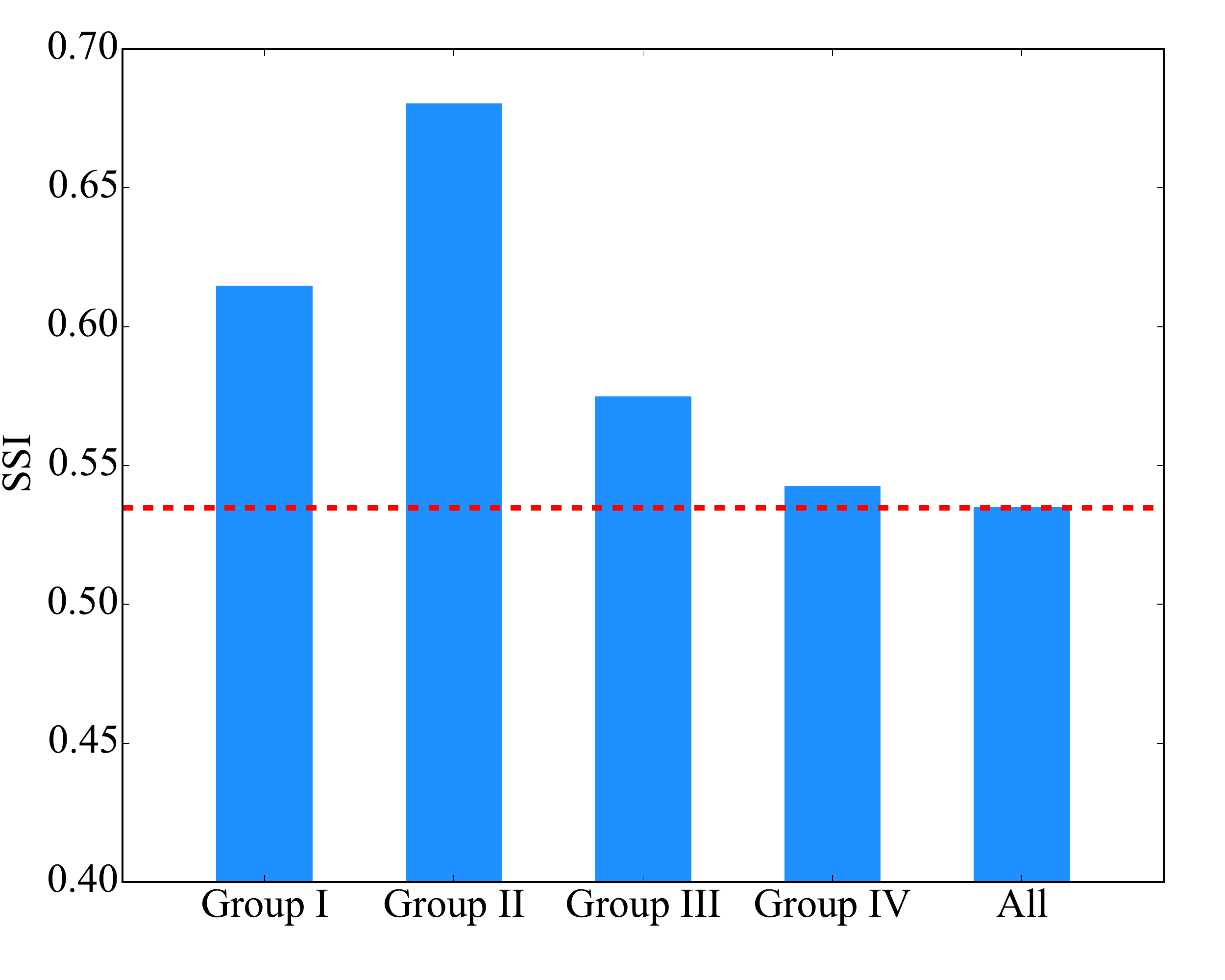}
\caption{
{\bf SSI of G-GM in grouped trajectories.} The dashed line represents the SSI for all the trajectories.
}
\label{fig:ssi_index_group}
\end{figure}

\clearpage
\begin{figure}[h!]
\includegraphics[width=7in]{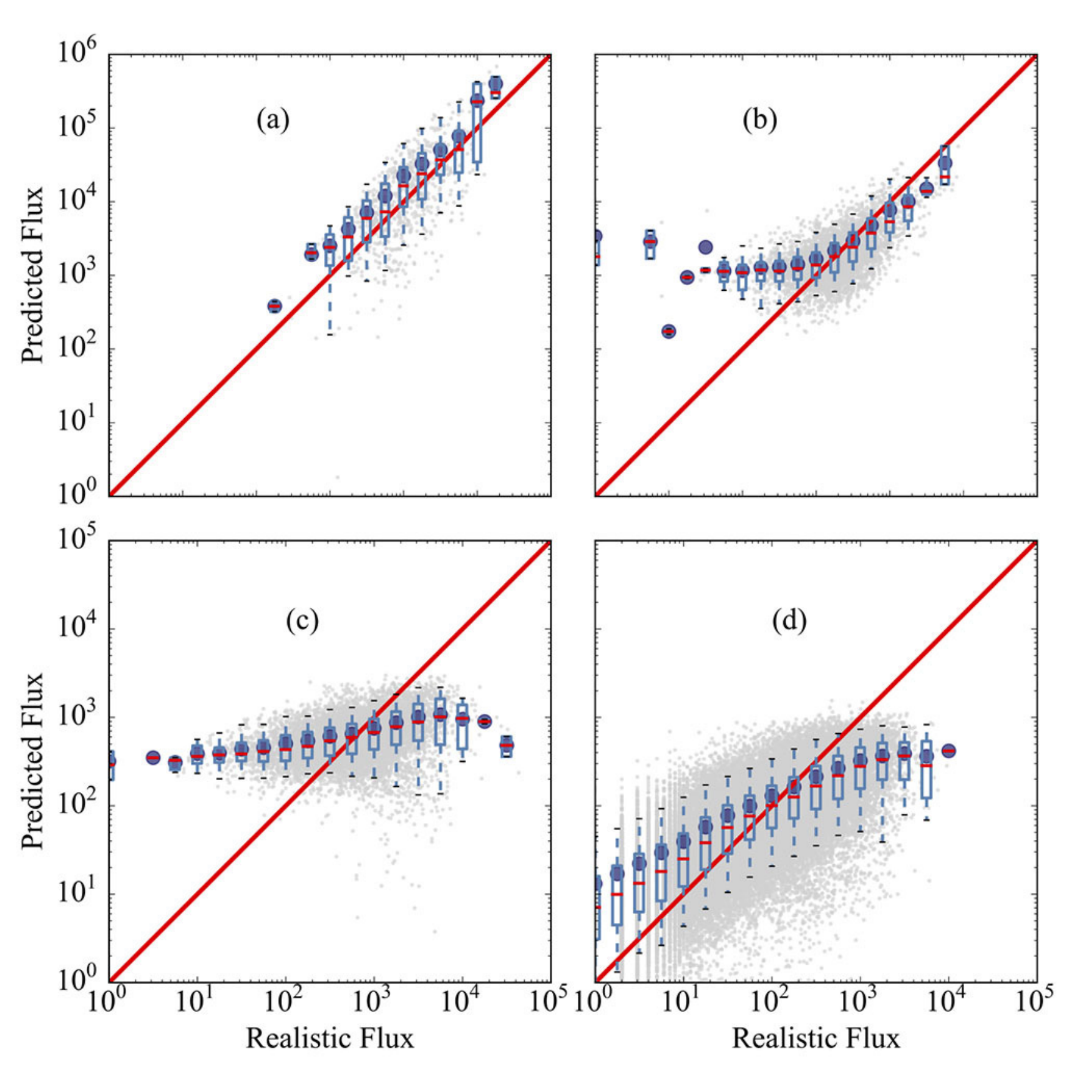}
\caption{
{\bf Modelling grouped trajectories.}
The comparison of the prediction capability of the grouped trajectories based on G-GM. (a) Pattern I; (b) Pattern II; (c) Pattern III; (d) Pattern IV.
}
\label{fig:model_group}
\end{figure}

\end{document}